\documentclass[12pt,preprint]{aastex}

\usepackage{color}
\usepackage{psfig}

\newcommand{\water}{\mbox{${\rm H}_{2}{\rm O}$}}
\newcommand{\app}{\ensuremath{\sim} }

\newcommand{\wwater}{\mbox{${\rm H}_{2}{\rm O}$}$\;$}

\shorttitle{Thermal Histories of Chondrules}
\shortauthors{Morris \& Desch}

\begin{document}

\title{Thermal Histories of Chondrules in Solar Nebula Shocks}

\author{M.~A.~Morris and S.~J.~Desch}
\affil{School of Earth and Space Exploration, Arizona State University,
        P.~O.~Box 871404, Tempe, AZ 85287-1404}

\email{melissa.a.morris@asu.edu}


\begin{abstract}
Chondrules are important early Solar System materials that can provide a wealth of information on conditions in the solar nebula, if their formation mechanism can be understood.
The theory most consistent with observational constraints, especially thermal histories, is the so-called ``shock model", in which chondrules were melted in solar nebula shocks.
However, several problems have been identified with previous shock models.  
These problems all pertained to the treatment of the radiation field, namely the input boundary condition to the radiation field, the proper treatment of the opacity of solids, and the proper treatment of molecular line cooling.
In this paper, we present the results of our updated shock model, which corrects for the problems listed above. 
Our new hydrodynamic shock code includes a complete treatment of molecular line cooling due to \water.
Previously, shock models including line cooling predicted chondrule cooling rates exceeding 10$^5$ K hr$^{-1}$.
Contrary to these expectations, we have found that the effect of line cooling is minimal; after the inclusion of line cooling, the cooling rates of chondrules are 10-1000 K hr$^{-1}$.
The reduction in the otherwise rapid cooling rates attributable to line cooling is due to a combination of factors, including buffering due to hydrogen recombination/dissociation, high column densities of water, and backwarming.
Our model demonstrates that the shock model for chondrule formation remains consistent with observational constraints.     

\end{abstract}


\section{Introduction}

The parent bodies of the most primitive meteorites, the chondrites, formed \app 4.57 billion years ago 
(Wadhwa \& Russell 2000).
Chondrites are remarkable for containing calcium-rich, aluminum-rich inclusions (CAIs), the oldest solids 
in the Solar System, whose formation has been dated to between 4567 Ma (Amelin et al.\ 2002, 2006; 
Jacobsen et al. 2008; Connelly et al. 2008) and 4569 Ma (Bouvier et al. 2007; Burkhardt et al. 2008; 
Bouvier \& Wadhwa 2009).
Also found in abundance within all chondrites (except for CI carbonaceous chondrites) are sub-millimeter- 
to millimeter-sized, (mostly ferromagnesian) igneous spheres, called chondrules, from which the chondrites 
derive their name. 
Chondrules formed, at most, \app~2-3 million years after CAIs (Amelin et al. 2002; Kita et al. 2005; 
Russell et al. 2006; Wadhwa et al. 2007; Connelly et al. 2008), as melt droplets that were heated to high 
temperatures while they were independent, free-floating objects in the early solar nebula (Lauretta et al.\ 2006).  
After they were heated, cooled, and crystallized, chondrules were incorporated into the parent bodies from which 
chondrites originate.  
Chondrules are capable of providing incredibly detailed information about conditions in the Solar System 
protoplanetary disk, if the process that led to their heating, melting and recrystallization could be understood 
(Connolly \& Desch 2004; Connolly et al.\ 2006; Lauretta et al.\ 2006).
Chondrules make up to 80\% of the volume of ordinary chondrites (Grossman 1988; Ciesla 2005; Lauretta et al. 2006), 
and it is estimated that \app 10$^{24}$ g of chondrules exist in the asteroid belt today (Levy 1988).
The asteroid belt has since been depleted by a factor of \app 1000 (Weidenschilling 1977, Morris \& Desch 2009).
indicating that there may have been \app 10$^{27}$ g of chondrules in the primordial belt (at least a Mars mass of rock). 
Such a prevalence of chondrules suggests that chondrule-forming events were widespread in the solar nebula.  
A process that can melt $10^{24} \, {\rm g}$ of rock is surely a dominant process in the solar nebula disk, 
and must be identified.

Any mechanism advanced to explain the melting of chondrules must meet the observational constraints on their 
formation, in particular, their thermal histories.  
We discuss, in detail, the constraints on chondrule formation in Section 2.
Proposed mechanisms for chondrule formation include interaction with the early active Sun, through jets 
(Liffman \& Brown 1995; Liffman \& Brown 1996) or solar flares at $<$ 0.1 AU (the so-called X-wind model of 
Shu et al.\ 1996, 1997, 2001; but see Desch et al., in prep), melting by lightning (Pilipp et al.\ 1998; Desch 
\& Cuzzi 2000), and crystallization melts produced by planetesimal impacts (Urey \& Craig 1953; Urey 1967; 
Sanders 1996; Lugmair \& Shukolyukov 2001).
The most widely-accepted hypothesis, though, is that chondrules were melted in shock waves in the protoplanetary 
disk (Hewins 1997; Jones et al.\ 2000; Connolly \& Desch 2004; Desch et al.\ 2005; Connolly et al.\ 2006),
driven either by X-ray flares (Nakamoto et al.\ 2005), gravitational instabilites (Boss 2001; Boss \& Durisen 
2005; Boley \& Durisen 2008), or planetesimal bow shocks (Hood 1998; Weidenschilling 1998; Ciesla et al. 2004; Nelson \& 
Ruffert 2005; Hood et al.\ 2005; Hood et al.\ 2009).
 
Passage through nebular shocks satisfies nearly all the experimental constraints on chondrule formation, not 
least among them the peak heating and two-stage cooling rates (Iida et al.\ 2001, hereafter INSN; Desch \& Connolly 
2002, hereafter DC02; Ciesla \& Hood 2002, hereafter CH02; Desch et al.\ 2005; Connolly et al.\ 2006; Krot 
et al.\ 2009).
Recent models are in good agreement with each other (Desch et al.\ 2005).  
For example, INSN found chondrule cooling rates $\sim 10^{4} \, {\rm K} \, {\rm hr}^{-1}$ in many cases, and 
DC02 and CH02 found cooling rates $\sim 10 - 1000 \, {\rm K} \, {\rm hr}^{-1}$.
Miura \& Nakamoto (2006) found that for plausible parameters, the cooling rates of chondrules and gas cluster 
around $5000 \, {\rm K} \, {\rm hr}^{-1}$.  
These values agree with each other and are distinct from the estimates of cooling rates predicted by other models, 
e.g., lightning (10$^5$-10$^6$ ${\rm K} \, {\rm hr}^{-1}$), and the X-wind model 
($< \, 10 {\rm K} \, {\rm hr}^{-1}$; Shu et al.\ 1996). 
For example, chondrules melted by lightning are not embedded in a hot gas, so will radiate their stored heat 
energy to free space.
The heat energy stored in a spherical chondrule of radius $a = 300 \, \mu{\rm m}$ 
and temperature $T = 2000 \, {\rm K}$ is $(4\pi/3)\rho a^3 \, C T$ = 9.7 x 10$^6$ erg,
where $\rho$ = 3.3 g cm$^{-3}$ is the density of silicate in chondrules
and $C$ = 1.3 x 10$^7$ erg g$^{-1}$ K$^{-1}$ is the specific heat capacity.
The rate at which the chondrule radiates away its energy is
$4 \pi a^2 Q \, \sigma T^4$ = 8.2 x 10$^6$ erg s$^{-1}$, where $q$ = 0.8 is the NIR emissivity
of silicate material (Li \& Greenberg 1997), and $\sigma$ is the Stefan-Boltzmann
constant.  Once heated, an isolated chondrule will cool in seconds,
whereas chondrule petrology argues for cooling over
tens of minutes.

While, overall, the nebular shock model successfully explains many aspects of chondrule formation, there
remain differences between the models (INSN; DC02; CH02) that lead them to infer different physical conditions
in the site of chondrule formation, as reviewed by Desch et al.\ (2005).
These model differences all involve the calculation of radiation effects, especially the radiative losses from 
molecular line emission, the opacity of solids, and the input radiation field.
DC02 and CH02 calculate the transfer of chondrule radiation, where INSN neglect this effect. 
The effect of line cooling---the cooling of gas by emission of so-called ``line photons" (photons of specific
wavelengths) by trace molecules in the gas such as CO and ${\rm H}_{2}{\rm O}$---in solar nebula shocks was 
considered by INSN and by Miura \& Nakamoto (2006). 
Line radiation from the water molecule ${\rm H}_{2}{\rm O}$, with its permanent electric dipole and its 
high cosmochemical abundance, is especially significant in a variety of astrophysical settings 
(e.g., Cernicharo \& Crovisier 2005).
INSN assumed a gas optically thin to the line radiation, whereas Miura \& Nakamoto (2006) allowed the gas to 
become marginally optically thick to this line radiation. 
DC02 and CH02 ignored the effect of line cooling, assuming an optically thick limit to the line radiation.
It remains to be resolved, and it is the point of the work presented here to resolve, which treatment is most realistic.

Another difference between the models is dust opacity.  
Only the opacity due to chondrules was considered by CH02, INSN and Miura \& Nakamoto (2006).  
In contrast, DC02 considered opacity due to both chondrules and a gray (no frequency or temperature dependence) opacity 
of micron-sized dust, $\kappa = 1.14 \, {\rm cm}^{2} \, {\rm g}^{-1}$, up to a dust evaporation temperature
$T_{\rm evap} = 2000 \, {\rm K}$, above which the dust opacity vanished. 

A final difference is the boundary condition assumed for the post-shock radiation field.
Both INSN and CH02 set the radiation field to be that of a blackbody with temperature 
$T_{\rm post} = T_{\rm pre}$, with little justification, while DC02 used the incorrect jump conditions 
of Hood \& Horanyi (1991) to derive a much higher post-shock temperature 
($T_{\rm post} \approx 1100 \, {\rm K}$ typically).
The isothermal assumption $T_{\rm post} = T_{\rm pre}$, strictly speaking, violates the assumption of 1-D, but
the jump conditions used by DC02 and Hood \& Horanyi (1991) were incorrect. 

The purpose of this paper is to properly include these physical effects (radiation boundary conditions, dust opacity 
and evaporation, and molecular line cooling), in order to better predict the thermal histories of chondrules.  
We describe our efforts to determine the appropriate input boundary conditions and the proper dust opacity and evaporation.  
We also describe our inclusion of molecular line cooling in the model and our assessment of its effects on the thermal 
histories of chondrules. 
In Section 2, we discuss the observational constraints on chondrule formation.  
In Section 3, we review the model and shock code of DC02 and describe our updates made in order to account for the
physical effects outlined above.  
In Section 4, we present our results and compare them with the results of DC02 and the meteoritic data.
In Section 5, we discuss the net effect of line cooling, in particular, backwarming and the buffering effects of hydrogen dissociation and 
recombination.
Finally, in Section 6, we discuss our conclusions and future work.   
   
\section{Constraints on Chondrule Formation} 

The textures and chemistry of chondrules can constrain their thermal histories, melt evolution, and precursor materials 
(Connolly et al.\ 2006).  
Experimental petrology determines the heating and cooling rates of chondrules by defining the constraints on crystal growth 
and evolution of the bulk composition during melting and cooling. 
Experimental petrology has shown that the most important determinants of texture during chondrule formation are peak 
temperature, cooling rates, and the presence of external seed nuclei.  
Cooling rates are constrained by texture, major and minor element abundances, and bulk chemistry (Connolly \& Desch 2004; 
Connolly et al.\ 2006; Lauretta et al.\ 2006).  
Peak temperature is constrained by the number of nuclei remaining in the melt and/or the number of nuclei encountered as 
external seed nuclei (Lofgren 1983, 1989, 1996; Hewins \& Connolly 1996; Hewins 1997; Connolly et al.\ 1998; Desch \& Connolly 
2002; Lauretta et al.\ 2006).      

\subsection{Thermal Histories of (Fe-, Mg-rich) Chondrules}

According to furnace experiments, in which melt droplets with chondrule compositions are allowed to cool and crystallize, 
reproduction of chondrule textures requires specific ranges of cooling rates (see below) between the liquidus temperature 
($\approx 1800$ K) and solidus temperature ($\approx 1400$ K; Hewins \& Connolly 1996).  
The majority of chondrules experienced peak temperatures in the range of 1770 - 2120 K for several seconds to minutes 
(Lofgren \& Lanier 1990; Radomsky \& Hewins 1990; Hewins \& Connolly 1996; Lofgren 1996; Hewins 1997; Connolly \& Love 1998;  
Jones et al.\ 2000; Connolly \& Desch 2004; Hewins et al.\ 2005; Ciesla 2005; Connolly et al.\ 2006; Lauretta et al.\ 2006), 
although the peak temperatures of barred olivine (BO) chondrules may have been as much as 2200 K (Connolly et al.\ 1998; 
Connolly et al.\ 2006).
Chondrule textures (the arrangement, shape, and size of their crystals) and the elemental zoning behavior within 
individual crystals constrain the cooling rates of chondrules (Connolly \& Hewins 1991; Jones \& Lofgren 1993; DeHart \& Lofgren 
1996; Desch \& Connolly 2002; Connolly et al.\ 2006; Lauretta et al.\ 2006).
Based on texture and chemistry, chondrules experienced cooling rates of $10 - 3000 \, {\rm K} \, {\rm hr}^{-1}$, with 
most cooling at \app 100 K hr$^{-1}$ or less through their crystallization range (Desch \& Connolly 2002; Hewins et al.\ 2005;
Connolly et al.\ 2006; Lauretta et al.\ 2006).  
Miyamoto et al.\ (2009) recently developed a model to calculate cooling rates using the Fe-Mg chemical zoning profiles 
of olivine.  
They found that for the type II porphyritic olivine chondrules in Semarkona, the cooling rates through crystallization 
temperatures are broadly consistent with furnace experiments (10 - 1000 K hr$^{-1}$).  
Initial cooling above the liquidus was at least 5000 K hr$^{-1}$ (Yu et al.\ 1996; Yu \& Hewins 1998; Desch \& Connolly 2002). 
Porphyritic chondrules cooled at about $10 -10^3 \, {\rm K} \, {\rm hr}^{-1}$, and barred-olivine chondrules cooled at about 
$10^3 \, {\rm K} \, {\rm hr}^{-1}$ (Hewins et al.\ 2005; see also Desch \& Connolly 2002 and references therein).
Additionally, chondrules retain volatile elements such as S, indicating that they did not remain above the liquidus 
for more than minutes, and cooled quite rapidly ($\gtrsim 10^4 \, {\rm K} \, {\rm hr}^{-1}$ while above the liquidus;
Yu \& Hewins 1998).
The presence of primary S tells us that chondrules did not experience prolonged heating between \app 650 - 1200 K for more 
than several minutes (Hewins et al.\ 1996; Connolly \& Love 1998; Jones et al.\ 2000; Lauretta et al.\ 2001; Tachibana 
\& Huss 2005; Connolly et al.\ 2006).
Finally, there is no indication of the isotopic fractionation that would arise from the free evaporation of alkalis 
such as Na, which constrains the time spent at high temperature before melting (Tachibana et al.\ 2004). 
Modeling of isotopic fractionation has shown that chondrules must heat up from 1300 to 1600 K in times 
on the order of minutes or less in order to prevent isotopic fractionation of S (Tachibana \& Huss 2005).  
The only alternative to such rapid heating is if the ambient nebular gas was considerably enriched in volatile or moderately 
volatile elements (Connolly et al.\ 2006).  

The time spent at the peak temperature also affects the amount of relict material that remains intact (Lofgren 1996; 
Connolly \& Desch 2004; Hewins et al.\ 2005; Connolly et al.\ 2006; Lauretta et al.\ 2006).  
Approximately 15\% of chondrules in ordinary chondrites contain relict grains (Jones 1996), the presence of which limits 
the duration of heating above the liquidus to tens of seconds to several minutes (Connolly et al.\ 2006).
The texture and chemistry of these relict grains indicates that they are previous generations of chondrules, signifying 
that chondrules experienced multiple heating events (Jones et al.\ 2005; Connolly et al.\ 2006; Lauretta et al.\ 2006).  
Evidence for multiple heating events is also found by the presence of fine-grained, igneous rims around some chondrules; 
a layer of material that was heated and melted in an event that post-dated melting of the host chondrule (Hewins et al.\
1996; Jones et al.\ 2005). 

\subsection{Additional Constraints on Chondrule Formation}

The combination of chondrules and the fine-grained matrix in chondritic meteorites results in a bulk composition very 
close to solar abundances, suggesting that the chondrules and matrix formed in the same vicinity within the solar nebula 
(Palme et al. 1993; Klerner \& Palme 2000; Scott \& Krott 2005; Ebel et al.\ 2008; Hezel \& Palme 2008).
Klerner \& Palme (2000) found a sub-chondritic value for the Ti/Al ratio in the matrix of the CR chondrite, Renazzo, 
while the Ti/Al ratio in the chondrules in it are supra-chondritic.  
The same complementarity of values for Ti/Al are found in Al Rais and the CV meteorites Kaba, Leoville, Mokoia, and 
Vigarano (Klerner \& Palme 2000).  
Additionally, the Mg/Si ratio in Renazzo matrix was found to be sub-chondritic, while the ratio in the chondrules are 
supra-chondritic (Ebel et al. 2008).  
Hezel \& Palme (2008) analyzed the Ca/Al ratios in the matrix and chondrules of Allende and Y-86751.  
These two CV meteorites are almost identical in bulk composition and structure.  
Hezel \& Palme found the Ca/Al ratio in the matrix of Allende to be sub-chondritic and the ratio in the matrix to be 
super-chondritic, yet the opposite is found for Y-86751.  
Hezel \& Palme interpret this as ruling out redistribution of Ca during parent body alteration, and therefore is indicative 
of a redistribution of Ca and Al among chondrules and matrix grains before accretion onto the parent body.
As these examples show, chondrules and the matrix grains that surround them in the chondrite thus formed from the same 
starting material, in the same vicinity of the solar nebula.
This must be considered as an additional constraint on chondrule formation models.  

Another constraint is the density of chondrules in the chondrule-forming region, which can be constrained in several
different ways.  
The frequency of compound chondrules, two or more chondrules that are fused together while molten, allows an estimate 
of the density of chondrules during formation of 1-10 m$^{-3}$ (Gooding \& Keil 1981; Wasson et al.\ 1995; Hood \& Kring 
1996; Ciesla et al.\ 2004a; Ciesla 2005; Yasuda \& Nakamoto 2008).
It is also worth noting that some compound chondrules---enveloping compound chondrules---provide evidence for multiple 
heating events (Wasson et al.\ 1995; Jones et al.\ 2005).  
Chondrule densities of \app 10 m$^{-3}$, over regions of \app 10$^3$ km, are also inferred from the lack of isotopic 
fractionation of volatiles during melting (Cuzzi \& Alexander 2006).  
Additionally, the retention of volatiles, such as Na, places rather firm lower limits on chondrule density of this order
as well (Cuzzi \& Alexander 2006; Fedkin et al.\ 2006; Alexander et al.\ 2008, Kropf \& Pack 2008).     

Finally, under the assumption that $^{26}$Al was homogeneous in the early Solar System, there is an apparent age gap 
between CAI and chondrule formation based on initial values of $^{26}$Al/$^{27}$Al (Russell et al.\ 1997; Galy et al.\ 2000; 
Tachibana et al.\ 2003; Bizzarro et al.\ 2004; Russell et al.\ 2007).
These same data suggest timescales for chondrule formation of several Myr (Huss et al.\ 2001; Tachibana et al.\ 2003; Wadhwa 
et al.\ 2007; Rudraswami et al.\ 2008; Hutcheon et al.\ 2009), as do Pb-Pb ages (Amelin et al.\ 2002; Kita et al.\ 2005; 
Russell et al.\ 2006; Connelly et al.\ 2008).

\section{The Shock Model}

\subsection{DC02 Shock Model}

The shock model used here builds directly on the model by DC02, which assumes a 1-D steady-state flow, conserving 
mass, energy, and momentum.  
We summarize the model below and refer the reader to DC02 for further details.
  
Any shock generated in the solar nebula will heat solids in three ways: 
1) by thermal exchange between the hot, dense, post-shock gas and the particles in the post-shock region;
2) by frictional heating, as the particles are slowed to the reduced post-shock velocity in the post-shock region; and
3) by absorption of infrared radiation emitted by heated particles everywhere, in both the pre-shock and post-shock regions.    
Shocks are manifestly transient, non-equilibrium structures. 
Shock models of chondrule formation must therefore account for the dynamics and energetics of chondrules and 
gas separately, including their interactions.
For typical parameters, such codes must also account for hydrogen dissociation and recombination.
They must also account for the transfer and absorption and emission of radiation. 
To minimize the complexity of the problem, DC02 considered a 1-D geometry, in which physical conditions were
assumed to vary only with distance $x$ from the shock front.
The calculations were also restricted to a range of $x$, the computational domain. 
Use of the 1-D approximation assumes implicitly that the lateral extent of the shock front greatly exceeds the 
computational domain.

The DC02 code assumes separate gas and solid fluids. 
The gas is divided into four populations: atomic hydrogen (H), molecular hydrogen (${\rm H}_{2}$), helium 
atoms (He), and molecules resulting from the evaporation of solids, which is represented with SiO.
Solids are divided into two types, dust grains and larger, spherical particles representative of chondrules.
Dust grains are assumed to be dynamically and thermally perfectly coupled to the gas, sharing its velocity
and temperature.
If at any point the dust grains exceed 2000 K they are assumed to be evaporated from that point forward. 
The chondrule-sized particles are divided into $J$ populations of identical particles, indexed by $j$.
Unique to each population of (spherical) particles are the number density $n_{j}$ of particles, their 
velocity $V_{j}$, their temperature $T_{j}$, and their radii $a_{j}$, as well as material properties.
All fluids are initialized at the pre-shock computational boundary with speed $V_{\rm s}$, with temperature
$T_{\rm pre}$. 
Their densities, velocities and temperatures are integrated forward using a fourth-order Runge-Kutta
integration, assuming a steady-state flow and applying the equations of continuity, a force equation 
including the drag force between gas and solids, and appropriate energy equations. 
The gas can absorb radiation energy (via the dust opacity), be heated by interaction with chondrules,
and can lose or gain heat energy through dissociations or recombinations of hydrogen molecules. 
Chondrules can emit or absorb radiation, or be heated by thermal exchange with the gas or by frictional 
heating.
 
Calculation of the radiation field followed the approach outlined in Mihalas (1978), assuming
plane-parallel, temperature-stratified slab atmospheres.
Given the incident radiation fields and the source function, $S(\tau)$ at all optical 
depths, the mean intensity of radiation, $J(\tau)$ (integrated over 
wavelength) is given by:
\begin{equation}
J(\tau) = \frac{I_{\rm pre }}{2} {\rm E}_{2} ( \tau_{\rm m} - \tau )
         +\frac{I_{\rm post}}{2} {\rm E}_{2} ( \tau ) + \frac{1}{2}
         \int_{0}^{\tau_{\rm m}} S(t) {\rm E}_{1} \left| t - 
         \tau \right| \, {\rm d} t,
\end{equation}
where $\tau$ is the optical depth, and ${\rm E}_{1}$ and ${\rm E}_{2}$ are the exponential integrals.
During the integrations, the radiation field is considered fixed while the gas and particle dynamics 
and energetics are calculated as a function of $x$.
It is assumed that $I_{\rm pre} = \sigma T_{0}^{4}$ in the pre-shock region, and
$I_{\rm post} = \sigma T_{\rm post}^{4}$ in the post-shock region.  
The calculation of $T_{\rm post}$, the post-shock temperature far from the shock front, is
described in \S 3.2.1.
After all other variables have been integrated across the entire computational domain, the radiation 
field is recalculated based on the updated particle densities, radii and temperatures.
The particles and gas are sent through the shock again, their temperatures calculated based on the 
new radiation field. 
The solution is iterated until the radiation field and the particle temperatures are self-consistent.
In this respect, the calculation is a so-called Lambda Iteration, which is known to converge slowly 
(Mihalas 1978). 
Indeed, in our typical runs, convergence of particle temperatures everywhere to within 1 K and 
convergence in $J_{r}$ to within $0.1\%$ were only achieved in about several hundred iterations, a 
number comparable to the optical depths across the computational domain.  
A more sophisticated treatment of the radiative transfer is, however, difficult to implement, and is 
subject to many feedbacks.
We chose to use the Lambda Iteration method for its ease of implementation, and for 
its demonstrated stability and convergence in reasonable computing times (hours).

The reader is referred to Appendix B and DC02 for further details of the original shock model. 

\subsection{Updated Shock Model}

We have modified the DC02 shock code to remedy the problems identified by Desch et al.\ (2005).
In this section we describe the updates to the radiation field jump conditions; we derive the
proper 1-D shock jump conditions for a gas-solids mixture, but show that it is usually justified
to assume that the post-shock gas returns to the pre-shock temperature.
We also discuss the appropriate dust opacity, and calculate the temperature at which dust
effectively evaporates.  
We show that 1500 K is a much better approximation than the 2000 K assumed by DC02. 
Most importantly, we include the emission, absorption and transfer of line radiation emitted by 
\wwater molecules.

\subsubsection{Jump Conditions Far from the Shock}

A major input to calculations of the radiation field (more specifically, the frequency-integrated
mean intensity, $J$, at all locations) is the input radiation field at the boundaries of the 
computational domain over which chondrule thermal histories are investigated, $I_{\rm pre}$ and
$I_{\rm post}$. 
Far from the shock front, in the pre-shock region, the radiation field $I_{\rm pre}$ is set to a 
blackbody radiation field at the ambient temperature, $T_{\rm pre}$, of the gas; but it is not 
immediately clear what the temperature, $T_{\rm post}$, in the post-shock region should approach.
Both INSN and CH02 set $T_{\rm post} = T_{\rm pre}$, with little justification, while DC02 used the 
incorrect jump conditions of Hood \& Horanyi (1991) to derive a much higher post-shock temperature 
(typically, $T_{\rm post} \approx 1100 \, {\rm K}$).
It is important to physically justify the isothermal assumption, $T_{\rm post} = T_{\rm pre}$, 
because formally it violates the assumption of 1-D, since the total energy of the system is not 
conserved; radiation in the direction lateral to the shock is required to carry away the energy 
in order for $T_{\rm post}$ to fall to $T_{\rm pre}$.
Although the jump conditions used by DC02 and Hood \& Horanyi (1991) did not make this assumption 
(they had $T_{\rm post} > T_{\rm pre}$), these jump conditions were incorrect because besides 
failing to consider the energy carried by solids, they used an incorrect expression for the 
radiative flux.  
Below we describe the proper jump conditions, then discuss why the isothermal assumption 
is probably justified after all.

Jump conditions relate physical conditions (e.g., density $\rho$, pressure $P$, temperature $T$,
velocity $V$) at a point before the shock to those after the shock (see Mihalas \& Mihalas 1984).
Immediately before and after the shock (i.e., a few meters), the jump conditions are those
of an ``adiabatic" shock because insignificant energy is radiated in that interval. 
The compression of the gas is then $\rho_{2} / \rho_{1} = \eta_{\rm AD}^{-1}$, where 
\begin{equation}
\eta_{\rm AD} = \frac{ 2 \gamma }{ \gamma + 1 } \, \frac{1}{ \gamma {\rm M}^{2} } + 
                \frac{ \gamma-1 }{ \gamma+1 }
\end{equation}
where $\gamma$ is the ratio of specific heats (Mihalas \& Mihalas 1984).  This compression term does not include solids, accounting for the gas only.
To include the heat energy of solids, we replace $\gamma$ with $\gamma'$, where
\begin{equation}
\gamma' = \frac{ \gamma + \delta (\gamma -1) }{ 1 + \delta (\gamma-1) },
\end{equation} 
and $\delta = (\rho_{\rm c} C_{\rm P} T_{\rm c})_{1} / P_{1}$, 
with $\rho_{\rm c}$, $C_{\rm P}$ and $T_{\rm c}$ referring to the chondrule density, heat capacity
and temperature, and the subscript 1 again referring to the pre-shock region.
In the limit that most heat is carried by solids (which do not experience a pressure force), 
$\delta \gg 1$, $\gamma' \approx 1$, and the system is effectively isothermal. 
For typical values in our canonical shock (\S 4.1), $\gamma = 1.49$ and $\gamma' = 1.30$.
Neglecting solids, we find $\eta_{\rm AD} \approx 0.20$ for strong shocks (${\rm M}^{2} \gg 1$),
and the gas is compressed by a factor $\eta_{\rm AD}^{-1} \approx 5$.  
Including the solids, we find a new compression factor $\eta_{\rm AD}'$ (found by 
replacing $\gamma$ in equation 2 with $\gamma'$), which is typically 0.13 in a strong shock, 
implying an even greater total compression of the gas plus chondrules in
the post-shock region, by a factor $\approx 8$, ultimately. 
These estimates ignore radiative losses; 
if radiation carries energy away from the shock front, one can show that the gas will compress by a larger
factor $\rho_{2} / \rho_{1} = \eta^{-1}$, where now $\eta$ solves the quadratic equation 
\begin{equation} 
(1 - \eta) \, (\eta_{\rm AD}' - \eta) = 
\frac{ \gamma'-1 }{ \gamma'+1 } \, \frac{ F_{2} - F_{1} }{ \rho_{1} V_{1}^{3} / 2 }, 
\label{1}
\end{equation}
where $F_{2} > 0$ and $F_{1} < 0$ are the radiative fluxes in the post-shock and pre-shock regions.
Once $\eta$ is determined, the post-shock temperature is easily found: 
\begin{equation}
T_{\rm post} = T_{\rm pre} \, \eta \, [ 1 + \gamma {\rm M}^{2} (1 - \eta) ].
\label{2}
\end{equation}
(see Appendix A or Morris 2009, PhD Thesis, for a complete derivation).
This formula relates the final post-shock temperature to the initial temperature and other shock properties.

\begin{figure}[ht]
 \includegraphics[width=14cm]{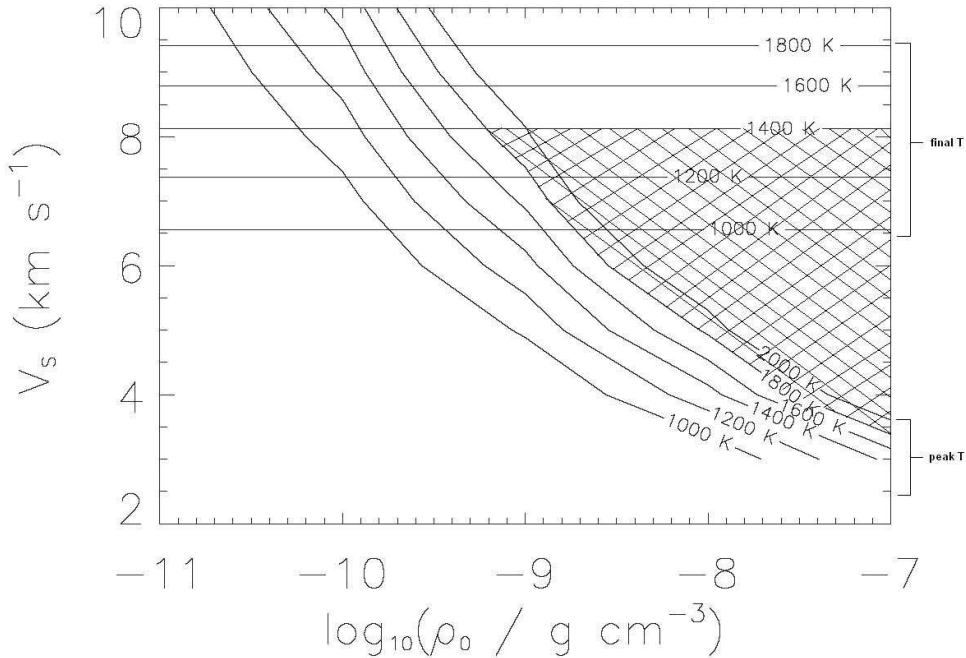}
\caption{Combinations of initial gas density, $\rho_0$, and shock velocity, $V_s$, potentially consistent with chondrule
thermal histories, assuming a 1-D shock.  The cross-hatched region delimits cases with $T_{\rm peak} > 1800$ K and 
$T_{\rm post} < 1400$ K, consistent with chondrule formation.  Only low shock speeds and gas densities higher than
expected are consistent with chondrule formation, if the 1-D approximation is strictly imposed.  }
\label{fig:jump}
\end{figure}

Using this proper jump condition for the temperature in the 1-D approximation, we have calculated the final
post-shock temperature for a range of initial densities and shock velocities.
The final temperature always exceeds the initial temperature, and can in fact be quite large: for 
$T_{\rm pre} = 300$ K, $T_{\rm post} > 1300 \, {\rm K}$ is typical unless the shock speed is small.
The {\it final} temperatures of chondrules melted in a 1-D shock will be $T_{\rm post}$.
We have calculated $T_{\rm post}$ for various combinations of ambient density, $\rho$, and shock 
speed, $V_{\rm s}$, as shown in Figure~\ref{fig:jump}.  
For the purposes of constraining the combinations of $\rho$ and $V_{\rm s}$ appropriate for 
chondrule-forming shocks, we also estimate the {\it peak} temperatures of chondrules.
Immediately after passing through the shock front, chondrules heat at a rate (per surface area) 
exceeding $\rho_{\rm g} \, V_{\rm rel}^{3} / 8$ and cool at rate no greater than $\sigma T^{4}$,
so a reasonable {\it estimate} is $T_{\rm peak} < (\rho_{\rm g} V_{\rm rel}^{3} / 8 \sigma)^{1/4}$, 
where $V_{\rm rel} \approx 5 V_{\rm s} / 6$ is the relative velocity between gas and chondrules. 
(Note that this estimate is only an approximation: when $\rho > 10^{-9} \, {\rm g} \, {\rm cm}^{-3}$ 
or so, for example, thermal exchange with the gas can cool chondrules effectively.) 
To be consistent with constraints on chondrule formation, a 1-D shock must yield $T_{\rm peak} > 1800$ K 
and $T_{\rm post} < 1400$ K.  
As illustrated in Figure~\ref{fig:jump}, this occurs only in cases of high density and low initial shock velocity 
($\rho_{\rm g} > 10^{-9} \; \rm g \; \rm cm^{-3}$ and $V_s \sim 4-7 \; \rm km \; \rm s^{-1}$).
While low shock speeds are plausible, these very high densities are not thought to have existed in the solar 
nebula at a few AU, where chondrules are presumed to form.
For example, even in a disk roughly 10 times more massive than the minimum mass solar nebula of 
Weidenschilling (1977), the densities do not exceed $\sim 10^{-9} \, {\rm g} \, {\rm cm}^{-3}$ at 2 AU 
(Desch 2007).
We conclude that for 1-D shocks in which chondrules reach temperatures high enough to melt, post-shock
temperatures will be too high to be consistent with chondrule thermal histories. 
Only by relaxing the 1-D assumption in some way can shocks be made consistent with chondrule formation. 

On the other hand, a truly 1-D shock should not, perhaps, be expected. 
One easily understood violation of the 1-D approximation involves radiative diffusion carrying energy away parallel to the shock 
front (for example, out the tops and bottoms of a disk in which the shock propagation direction lies in the plane of the disk).
The radiation generated by a nebular shock with lateral extent $L$ will diffuse on a timescale 
\begin{equation}
t_{\rm rd} = \frac{3 \rho_{\rm g} C_{\rm V} L^2 }{64 \pi^2 \, \lambda \, \sigma T^{3} }
\end{equation}
(Mihalas \& Mihalas 1984), where $C_{\rm V}$ is the heat capacity of the gas and $\lambda$ is the mean free path of photons. 
For our preferred post-shock dust opacity ($\kappa$ = 0.03 cm$^2$ g$^{-1}$), a post-shock density 
$\rho_{\rm g} = 6 \times 10^{-9} \, {\rm g} \, {\rm cm}^{-3}$ and temperature 2000 K, 
$t_{\rm rd} = 1.3 \times 10^{7} \, (L / H)^{2} \, {\rm s}$, where $H \approx 0.2 \, {\rm AU}$
is the scale height of the disk.  
For opacity due to a solar composition of $300 \, \mu{\rm m}$ chondrules, i.e., where dust evaporates, $t_{\rm rd}$ is
lowered by a factor of 200.
This is to be compared to the time for gas and chondrules to reach the computational
boundary in simulations, typically $(5 \times 10^{6} \, {\rm km}) / (1 \, {\rm km} \, {\rm s}^{-1})$
$\approx 5 \times 10^{6} \, {\rm s}$.
Thus, the time for radiation to diffuse a lateral distance $L \sim H$ is comparable to the dynamical timescales
within the shock, and the assumption that no radiation is lost from the shock region is invalid.
Thus the 1-D condition is violated. 
Especially since radiative diffusion is most rapid at high temperatures, and especially if dust evaporates
in the shock, it is reasonable to assume that gas cools before reaching the computational boundary in our models, 
effectively to $T_{\rm pre}$. 
We set $T_{\rm post} = T_{\rm pre}$ in the models that follow, acknowledging that a 2-D radiative transfer 
calculation is required to truly capture the exact radiative losses in these shocks. 

\subsubsection{Dust Opacities}

Previous shock models have crudely estimated dust evaporation temperatures and dust opacities, where 
dust opacity was included. 
CH02, INSN and Miura \& Nakamoto (2006) only considered the opacity due to chondrules, not dust.
In contrast, DC02 considered opacity due to both chondrules and a gray (no frequency / temperature dependence) 
opacity of micron-sized dust, with $\kappa = 1.14 \, {\rm cm}^{2} \, {\rm g}^{-1}$ (per gram of gas), up to
a dust evaporation temperature $T_{\rm evap} = 2000 \, {\rm K}$, above which the dust opacity was assumed
to vanish.
For meteoritic abundances of micron-sized dust and chondrules, the opacity from dust dominates over that 
from chondrules, so it is important to include dust opacity, especially as such opacity will be important 
in shutting off cooling by line emission (Morris et al. 2009).  
Dust opacities are wavelength-dependent. 
Assuming a dust-to-gas ratio $\rho_{\rm d} / \rho_{\rm g} = 5 \times 10^{-3}$, a particle radius 
$a_p = 0.5 \, \mu{\rm m}$, and an absorptivity $Q_{\rm abs} = 1$ for $\lambda < 2\pi a_s$ and 
$Q_{\rm abs} = 2\pi a_s/\lambda$ for $\lambda > 2\pi a_s$, we derive 
\begin{equation}
\kappa_{\lambda} = 30 \, \min \left[ 1, (\lambda / 3.1 \, \mu{\rm m})^{-1} \right] \, {\rm cm}^{2} \, {\rm g}^{-1}. 
\label{eq:ourkappa1}
\end{equation}
Our estimates are similar to those derived by Henning \& Stognienko (1996), who used a particle size distribution 
instead of our simplified monodispersion.  
Ideally, the calculation of the radiation field should be performed at multiple wavelengths to account for the 
wavelength-dependence of the opacity; in practice, however, this is computationally expensive, and we calculate
only a wavelength-integrated radiation field. 
Using a set of just 30 frequencies would increase our computational time for each simulation from \app 10 days to $>$ 30 days.  
We therefore desire a wavelength-integrated approximation to the dust opacity, as well. 

We use the Planck mean opacity and the Rosseland mean opacity as starting points in our search for such an approximation.  
In stellar atmospheres (and other opaque media) the Rosseland mean intensity is related to the mean free path of the dominant photons, and
is used to express the radiative energy flux using the diffusion approximation.
The Planck mean opacity, in contrast, appears naturally
when integrating the radiative transfer equation over frequency.  
As we have calculated the radiative energy transfer based on the radiative transfer equation, 
we have used the Planck opacity
as our wavelength-averaged opacity. 
From its definition, 
\begin{equation}
\kappa_{\rm P}(T) = \frac{ \int_{0}^{\infty} \, \kappa_{\lambda} \, B_{\lambda}(T) \, d\lambda } 
                      { \int_{0}^{\infty} \, B_{\lambda}(T) \, d\lambda },
\end{equation} 
the Planck opacity is seen to be temperature-dependent. 
Integrating the opacity $\kappa_{\lambda}$ above over wavelength, we find an excellent approximation for $\kappa_{\rm P}(T)$, over temperatures ranging from the ambient temperature up to the evaporation temperature (300 - 1500 K). 
For a solar composition ($\rho_{\rm d} / \rho_{\rm g} = 5 \times 10^{-3}$), our approximation to the opacity is
\begin{equation} 
\kappa_{\rm{app}}= 12.161 \; \ln(T / 1 \, {\rm K}) - 62.524 \; {\rm cm}^{2} \; {\rm g}^{-1}\, 
\label{appkappa}
\end{equation} 
per gram of gas.  
Calculations of the optical depth employ this opacity.

\subsubsection{Dust Evaporation}

Dust grains will typically evaporate in chondrule-forming shocks (DC02; Wasson 2008), affecting the opacity of 
the gas.
If they do not do so before the shock front, dust grains are very likely to evaporate at the shock front itself.
Unlike chondrules, which take minutes to slow (during which time they can radiate), micron-sized dust grains slow 
past the shock in only milliseconds and are very poor radiators (being smaller than the wavelength of maximum emission).
Their kinetic energy is converted predominantly into heat within the dust grains. 
Because the specific kinetic energy ($V^2 / 2$) of dust grains exceeds their latent heat of evaporation 
($l_{\rm evap} \sim 10^{11} \, {\rm erg} \, {\rm g}^{-1}$) when $V > 5 \, {\rm km} \, {\rm s}^{-1}$, 
dust grains are likely to evaporate in milliseconds in chondrule-forming shocks. 
DC02 assumed temperatures $> 2000 \, {\rm K}$ were required to evaporate dust and found that dust grains usually
did evaporate at the shock front.
Temperatures this high were not obtained in the pre-shock region, although DC02 did find that dust grains are 
heated to high temperatures $> 1000 \, {\rm K}$ before the shock, by absorption of radiation from the post-shock 
chondrules.
Temperatures of 2000 K are much higher than the typical evaporation temperatures of silicates and metals 
($\approx 1350 \, {\rm K}$; Lodders 2003), however, and evaporation in the pre-shock region is not precluded. 
Here we investigate the ability of dust grains to evaporate in the pre-shock region as well. 

First, we calculate the rate at which dust grains evaporate. 
Richter et al. (2002) provide the following temperature-dependent evaporation rate for materials: 
\begin{equation} 
J_i= \sum_{j=1}^n \; \frac{n_{ij}\gamma_{ij}P_{ij}^{sat}}{\sqrt{2 \pi m_{ij}RT}},
\label{eq:evap}
\end{equation} 
in units of mol cm$^{-2}$ s$^{-1}$, where $i$ is the isotope or element considered, $j$ is the gas species 
containing $i$, $n$ is the number density of $i$, $\gamma$ is the evaporation coefficient of $i$, $P^{sat}$ is 
the saturation 
vapor pressure for $j$, $m$ is the molecular weight of $j$, $R$ is the gas constant, and $T$ is the temperature.
Davis \& Richter (2005) give the temperature-dependent evaporation coefficients for forsterite, and calculate 
the vacuum evaporation rate as a function of temperature, as well as the evaporation rate at 1773 K as a function 
of pressure.
We neglect the temperature dependence of the saturation vapor pressure (a small effect, at most a few times 
$10^{-2}$ g cm$^{-2}$ s$^{-1}$), and substitute the appropriate ratio $\gamma/T$ into the given evaporation rate 
at 1773 K to calculate the evaporation rates at other temperatures.  
At 10$^{-3}$ bar, we find a forsterite dust grain with radius $a = 0.5 \mu$m will evaporate in $<$ 60 s upon 
reaching a temperature $T \, \approx \, 1500$ K.  
By the time the dust has achieved a temperature of \app 1800 K, it will evaporate in $<$ 10 s.  

Within our code, dust grains are assumed to evaporate instantaneously at a particular temperature. 
In this context, evaporation can be considered instantaneous when dust grains evaporate before they 
move significantly. 
In the pre-shock region, the most significant physical effect of the dust grains is to provide optical depth. 
For typical pre-shock densities ($\rho_{\rm g} \approx 10^{-9} \, {\rm g} \, {\rm cm}^{-3}$) and opacities
($\kappa \approx 26 \, {\rm cm}^{2} \, {\rm g}^{-1}$ at 1500 K), an optical depth of unity is achieved over
lengthscales $\sim 400 \, {\rm km}$. 
Dust grains typically traverse these distances in $\sim 50 \, {\rm s}$, so dust grains can be considered to evaporate
instantaneously when their temperatures exceed roughly 1500 K. 
In the models that follow, we consider dust grains to be destroyed once their temperatures exceed 1500 K.
We note that this does not eliminate dust opacity entirely, however, as there will always be some opacity due to 
ultra-refractories such as Al (\app 0.001 of the original opacity; Lenzuni et al. 1995), as well as from the 
chondrules themselves.
We include the opacity of refractories at all temperatures, including temperatures
in excess of 1500 K, by taking 0.001 x the opacity of the dust, calculated as if it had not evaporated (see equation~\ref{appkappa}).

\subsubsection{Inclusion of Line Cooling in the Shock Code}

The last update we make to the DC02 code is the inclusion of line radiation, the absorption and emission
of line photons emitted by ${\rm H}_{2}{\rm O}$ molecules specifically.
We build on the work of Morris et al.\ (2009), who calculated the cooling rate of gas due to line radiation emitted 
by $\water$ molecule, at gas densities appropriate to protoplanetary disks.
The cooling of a slab of gas depends on the temperature and number density of water molecules in the slab, as well
as the column density of water between the slab and the external medium to which the line photons must escape. 
These rates were calculated by Morris et al.\ (2009) and applied to a toy model of a chondrule-forming shock.
This toy model assumed that chondrules had the same temperature as the gas, and that the gas cooled past the
shock front by emitting line photons to the cooler, pre-shock region.
The toy model did not include the compression of the gas in the post-shock region, or absorption and emission
of infrared radiation, and therefore yields only an approximation to the post-shock conditions.
In addition, hydrogen dissociation and recombination were neglected.
Nonetheless, the toy model provided the useful insight that the gas potentially could cool considerably due to 
line emission before the gas moves significantly far from the shock front.
(In the toy model, line cooling effectively shuts off only when either the column density of water or the optical 
depth of solids, between the gas and the shock front, becomes large enough.)
Morris et al.\ (2009) found that for typical parameters, without dust, $\water$ line cooling could reduce the 
temperature of the gas to 1400 - 1800$\;$K in $\leq$ 0.1 hr; even then, the column density of water is low enough
that cooling continues at this rate ($>$10$^4$ K/hr), as in the model of INSN. 
With the inclusion of dust, however, the grains absorb line photons and inhibit cooling; the gas cools several 
hundred K in the first $\sim 0.1$ hr, but ceases to cool thereafter.
Chondrules would be inferred to have a very rapid cooling rate initially, $\sim 10^4 \ {\rm K} \, {\rm hr}^{-1}$, 
dropping several hundred K within $\sim 0.1$ hr, then cooling thereafter at a much slower rate, if dust did
not evaporate in the post-shock region. 
The approximate toy model of Morris et al.\ (2009) identified several important factors and showed that line 
cooling could produce a measureable change in the thermal histories of chondrules; it therefore motivated the 
present study to include line cooling and the transfer of line radiation in the full shock code. 

Morris et al.\ (2009) calculated the cooling rates of gas due to line emission by computing the escape probability
of photons produced by 1.2 million transitions of the \wwater molecule.
Reproducing these calculations is computationally prohibitive, and we seek instead an effective wavelength-integrated
cooling rate applicable for all column densities. 
This approximation must also account for the (wavelength-dependent) ability of dust grains and solids to 
absorb line photons. 
We have found that for all temperatures, in all cases of interest (low dust opacity), the following approximation 
provides an excellent (to within 10 \%) fit to the cooling rates: 
\begin{equation}
\Lambda(N_{\rm H_2O},\Sigma_{\rm eff})=\Lambda(N_{\rm H_2O},\Sigma_{\rm eff}=0) \; 
\exp\left[ - \frac{\Sigma_{\rm eff}}{1.2 \; \rm x \; 10^{-4} \; \rm g \; \rm cm^{-2}} \right],
\label{eq:coolapp}
\end{equation}
where $1.2 \; \rm x \; 10^{-4} \; \rm g \; \rm cm^{-2}$ is the column density of solids that gives an effective 
wavelength-averaged optical depth of 1, and 
\begin{equation}
\Sigma_{\rm eff} = \Sigma_{\rm d} + 8 \; {\rm x } 10^{-4} \; \Sigma_{\rm ch},
\label{eq:effcolumn}
\end{equation}
where the column density of dust from the computational boundary to each zone $i$ is calculated as 
\begin{equation}
\Sigma_{{\rm d},i} = \sum_{j=2}^i \frac{1}{2} \left( \rho_{{\rm d},{j}} + \rho_{{\rm d},{j-1}} \right ) \; \left(x_j - x_{j-1}\right),
\label{eq:dustcol}
\end{equation}
(where $\Sigma_{{\rm d},1} \equiv 1$), 
with a similar expression for the column density of chondrules/chondrule precursors and for the column density of
water.
Utilizing the complete methods of Morris et al.\ (2009), lookup tables of exact cooling rates without dust, 
$\Lambda(N_{\rm H_2O},\Sigma_{\rm d}=0)$, were generated for 48 column densities of \wwater ranging between $10^{13}$ and 
$10^{25} \; {\rm cm}^{-2}$, and temperatures of 250 - 4000 K (in increments of 250 K), for a total of 720 entries in the 
$N_{\rm H_2O}$ - $T$ grid.  
This was used to calculate the rate at which line photons are emitted from any region in front of or behind the shock,
and escape to various distances. 

In terms of $\Lambda$, the flux of line photons emitted from a zone $i$ (in either direction) is 
$n_{\rm H_2O_{i}} \, \Lambda(N_{\rm H_2O},\Sigma_{\rm eff} = 0, T_{i}) \, dx_{i}$, (with units of
erg cm$^{-2}$ s$^{-1}$).
The fraction of this energy that is absorbed in a zone $j$ is the energy that escapes from $i$ all the way
to the near edge of zone $j$, minus the energy that escapes from $i$, out the far edge of zone $j$.
For example, if $j > i$, the energy flux emitted in zone $i$ and absorbed in zone $j$ is 
\[
\dot{E}_{i \rightarrow j} = \frac{1}{2} \; n_{\rm H_2O_i} \; \left[ \Lambda(N_{\rm H_2O_{j-1}}-N_{\rm H_2O_{i}},\Sigma_{{\rm eff}_{j-1}}-\Sigma_{{\rm eff}_{i}},T_i)
\right.
\]
\begin{equation}
\left.
- \Lambda(N_{\rm H_2O_{j+1}}-N_{\rm H_2O_{i}},\Sigma_{{\rm eff}_{j+1}}-\Sigma_{{\rm eff}_{i}},T_i) \right] \; dx_i.
\end{equation}
In the ``optically thin" limit, this is proportional to $N_{{\rm H2O},j+1} - N_{{\rm H2O},j}$, and the 
energy absorbed per unit volume in $j$ is proportional to $n_{\rm H2O}$ in that zone, as expected. 
(See Appendix C for exact expressions.)
Because $\Sigma_{\rm eff}$ can also affect how much radiation passes through zone $j$, absorption by solids
is also included in this treatment. 
After calculating $E_{i\rightarrow j}$ for all combinations of $i$ and $j$, 
the net heating rate per unit volume in zone $i$ is readily found:
\begin{equation}
\dot{e}_i = \frac{1}{dx_i}\; \sum_{j \neq i} \; \left( \dot{E}_{j\rightarrow i} - \dot{E}_{i\rightarrow j} \right).
\label{eq:edotnetmain}
\end{equation}
This net heating rate is assumed to apply to the gas, as an added term in the gas thermal evolution
equation (i.e., equation 23 of DC02). 

The approach we have adopted captures the essential physics of the problem but is not technically
complete in every detail. 
First, line photons can be absorbed by water molecules, dust grains or chondrules, but in all cases 
we consider the energy to heat the gas.
If photons are absorbed by dust grains, their energy is directly communicated to the gas by
effective thermal exchange between gas and dust grains.  
Chondrules could be heated directly by absorbing line photons, but in fact insignficant amounts of 
energy are exchanged between chondrules and water molecules via radiation (see Appendix C). 
Second, we ignore the ability of gas molecules to absorb continuum photons: where dust exists, 
gas will absorb continuum photons via dust; where it does not, gas will absorb only a small fraction 
(we estimate $\approx 2\%$, based on the mean free paths of {\it line} photons through the gas
and through chondrules), of the total radiation field, and leave the remaining 98\% untouched. 
The propagation of the continuum radiation field is not affected by neglecting absorption of
continuum photons by the gas. 
In all cases where we have made approximations, we have erred on the side of allowing the gas to
cool and decreasing the radiative communication between the gas and chondrules. 

One of the most important, physically relevant effects captured by our approach, not captured in 
previous attempts to include line cooling, is that the emission of radiation from zone $i$ to other 
zones $j$ is always accompanied by the absorption in zone $i$ of radiation from other zones $j$.
Even if $T_{i} > T_{j}$, some radiation from zone $j$ will be absorbed in zone $i$, limiting the 
net cooling in that zone, an effect we term ``backwarming".
Because of this effect, gas does not cool to an infinitely distant medium much colder than itself;
rather, it radiates to gas an average column density $N_{\rm H_2O} \sim 10^{19} \, {\rm cm}^{-2}$ 
(typically 100 km) away, at roughly the same temperature as itself.  
This severely limits the ability of gas to cool by emitting line radiation. 
Our updated shock code now includes all the improvements suggested in Desch et al. (2005);
using it, we were able to evaluate the effects of molecular line cooling.

\section{Results}

\subsection{DC02 Results}

In order to illustrate the relevant physics in a chondrule-forming shock, and provide a baseline 
case to illustrate the effects of the physical updates we have made to the DC02 code, we first discuss
the output of the DC02 code itself, for what we term a canonical shock. 
The chondrule precursors are here assumed to be spherical, with radius $300 \, \mu{\rm m}$, and density 
$3.3 \, {\rm g} \, {\rm cm}^{-3}$. 
It is assumed that averaged over the nebula, the mass of solids is a fraction $0.005$ of the mass of gas, 
and that 75\% of the solids' mass is in the form of chondrules, the rest residing in micron-sized dust grains
(consistent with the proportions in ordinary chondrites).
The fraction of the gas mass that is chondrules is therefore $0.00375$ when averaged over the nebula.
In other runs, we increase or decrease the density of chondrules overrun by the shock by a ``concentration"
factor ${\cal C}$ relative to this fraction; in the DC02 canonical shock ${\cal C} = 1$. 
Consistent with models of the solar nebula (e.g., Desch 2007), we assume an ambient gas density 
$\rho_{\rm g} = 1 \times 10^{-9} \, {\rm cm}^{-3}$.
The speed of the shock is taken to be $7 \, {\rm km} \, {\rm s}^{-1}$. 
This speed, a fraction of the Keplerian velocity $\approx 20 \, {\rm km} \, {\rm s}^{-1}$ at 2 AU, is 
consistent with shocks driven by gravitational instabilities (Boss \& Durisen 2005). 

\begin{figure}[ht]
\includegraphics[width=14cm]{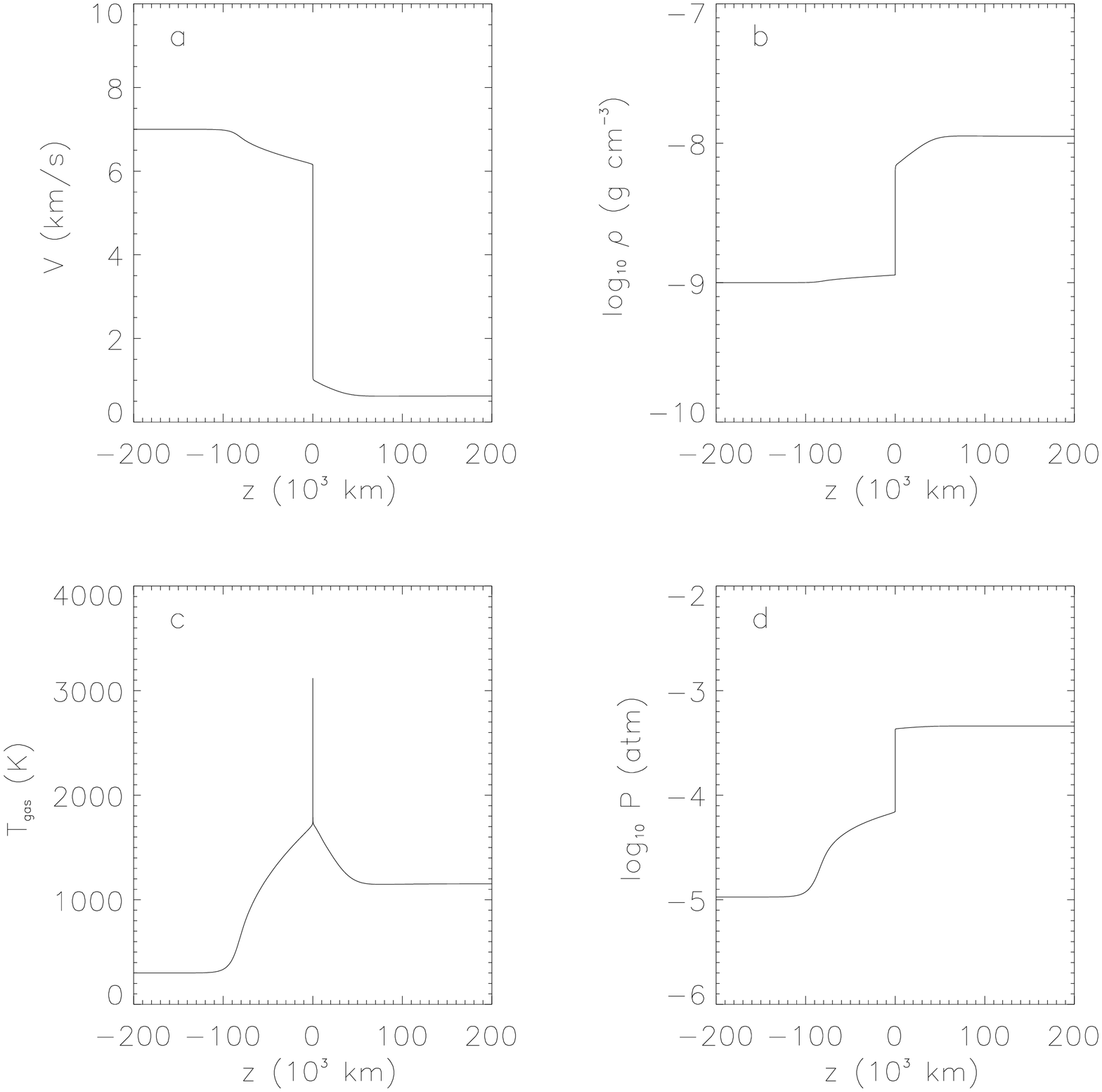}
\caption{Gas properties [(a) velocity, (b) density, (c) temperature, and (d) pressure] as a function of the distance 
{\it z} from the shock front.  The pre-shock region here is on the left.  A steady state flow is assumed; the properties 
of an individual gas parcel as a function of time are found by reading these graphs from left to right.  These results 
were calculated using the original code of DC02, with $V_s = 7 \, {\rm km} \, {\rm s}^{-1}$ and ${\cal C} = 1$.  (See text for additional shock parameters and details).}
\label{fig:DC02fig1}
\end{figure}

\begin{figure}[ht]
\includegraphics[width=14cm]{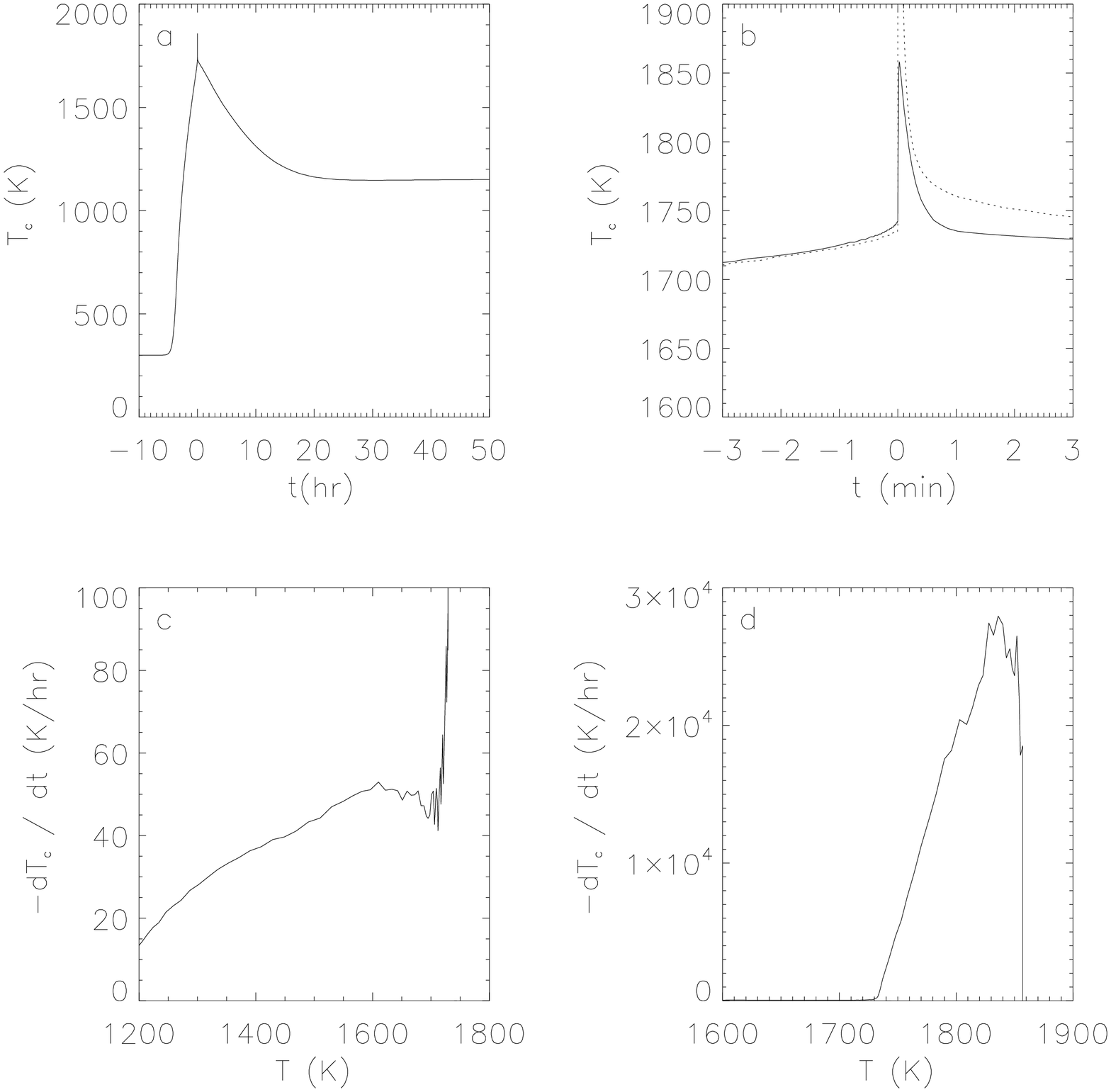}
\caption{Thermal histories of chondrules for the same shock as in Figure 1, over the course of hours (a), and also 
over minutes (b), where they are contrasted with the temperatures of the gas (dotted line).  The cooling rates of 
chondrules as a function of temperature through the crystallization temperatures (c), and at higher temperatures (d).  These results 
were calculated using the original code of DC02, with $V_s = 7 \, {\rm km} \, {\rm s}^{-1}$ and ${\cal C} = 1$.  (See text for additional shock parameters and details).}
\label{fig:DC02fig2}
\end{figure}

The variation of gas velocity, density, temperature and pressure as a function of distance $z$ from the shock 
front are displayed in Figure~\ref{fig:DC02fig1}. 
In the pre-shock region, the gas properties remain constant until within $\sim 10^{5} \, {\rm km}$ of the shock front.
At that point, dust and chondrules in the gas begin absorbing radiation from the other side of the shock front.
The hot chondrules transfer heat to the gas by thermal collisions, and the gas temperature and pressure increase.

In a steady-state 1-D shock, this pre-heating of the gas is inevitable, as radiation diffuses into the pre-shock 
region as a Marshak wave (Mihalas \& Mihalas 1984). 
The radiation diffuses a distance $L$ into the pre-shock region such that the radiative diffusion time $t_{\rm rd}$
based on $L$ (see equation 1) equals the dynamical time, the time needed to travel a distance $L$ to the shock front.
Because of the radiative diffusion, the increase in gas pressure exerts a backward force on the gas, and its velocity 
decreases and its density increases.
Before the gas even reaches the shock front, its velocity has slowed from $7 \, {\rm km} \, {\rm s}^{-1}$ to 
$6.15 \, {\rm km} \, {\rm s}^{-1}$, and its temperature has increased from $300 \, {\rm K}$ to $1736 \, {\rm K}$.
The decrease in velocity and the increase in the sound speed reduce the Mach number from 6.78 to only 2.48. 
If radiation were more effective, the diffusion of energy across the shock front could change the shock into a 
subsonic flow with no sharp boundary.
In the shock considered here, the flow remains supersonic, and the change in state of the gas is abrupt, although 
the impact across the shock front is lessened.
Despite this, the post-shock temperature is not much affected by the radiation diffusion, and is 3120 K right after 
the shock hits.
Soon after the shock hits, the hydrogen in the gas finds itself out of equilibrium at such a high temperature.
Dissociation of hydrogen is rapid, and each dissociation consumes 4.48 eV per hydrogen molecule.
The gas rapidly cools to temperatures closer to 2000 K. 
After a few minutes the gas and chondrules are in thermal equilibrium, and both cool only as fast as they can
move several optical depths from the shock front, where most of the radiation is produced.   

The thermal history of chondrules, including heating and cooling from interactions with the gas and radiation 
fields, is illustrated in Figure~\ref{fig:DC02fig2}. 
At very long times before the chondrules reach the shock front, they are in equilibrium with the gas, staying at an 
ambient temperature here assumed to be 300 K. 
Beginning about 4 hours (in this case) before the shock front overtakes the chondrules, they begin absorbing thermal 
radiation emitted by the chondrules that have already been through the shock front. 
This heats them up to 500 K at 230 minutes before they pass through the shock front, 1000 K at 170 minutes before 
the shock, and 1500 K at 64 minutes before the shock hits.
By the time they reach the shock front proper, the chondrules have already reached temperatures $\approx 1742 \, {\rm K}$.
After they pass through the shock front, they find themselves moving supersonically with respect to the gas, and gas-drag
frictional heating heats them to a peak temperature of 1858 K at $1.6$ seconds after the shock hits. 
After this peak, the temperature drops rapidly ($\sim 3 \times 10^{4} \, {\rm K} \, {\rm hr}^{-1}$) to a ``baseline''
temperature of $\sim 1700 \, {\rm K}$, where the heating is due to radiation and thermal exchange with the gas, but 
not gas-drag heating.
Once this baseline temperature is reached, the rate of change of the temperature is determined not by the particle 
dynamics, but by how quickly the particle can escape the zone of intense thermal radiation and hot gas. 
The cooling rates, in the range $35-50 \, {\rm K} \, {\rm hr}^{-1}$ over the temperature range
$1400 - 1700 \, {\rm K}$ at which significant crystallization will take place, are consistent with constraints on the thermal 
histories of chondrules, as is the initial rapid cooling rates, and the initial temperature.

\begin{figure}[ht]
\includegraphics[width=14cm]{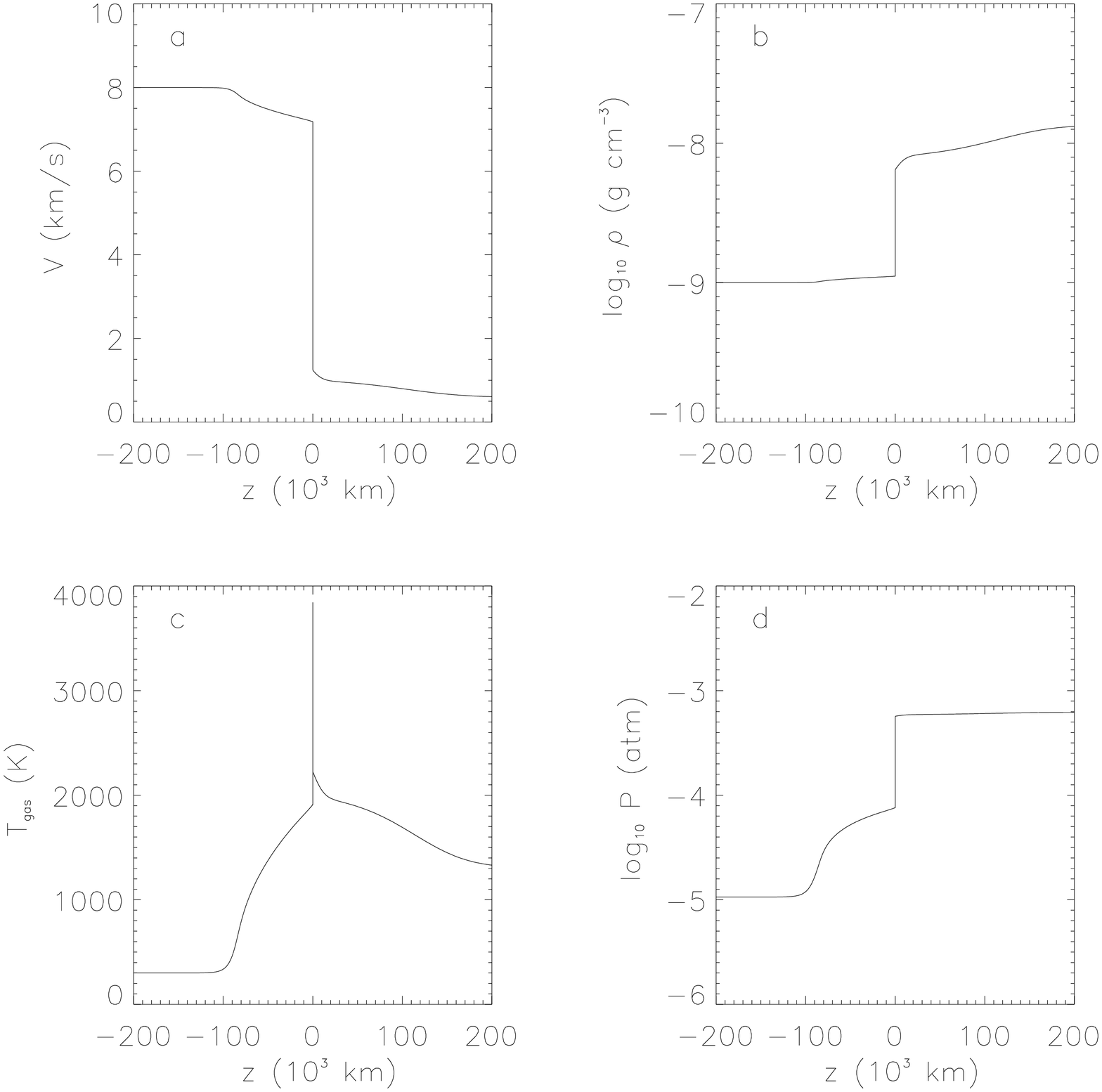}
\caption{Same as Figure~\ref{fig:DC02fig1}.  These results 
were calculated using the original code of DC02, but with $V_s = 8 \, {\rm km} \, {\rm s}^{-1}$ and ${\cal C} = 10$.  (See text for additional shock parameters and details).}
\label{fig:1SD8}
\end{figure}

\begin{figure}[ht]
\includegraphics[width=14cm]{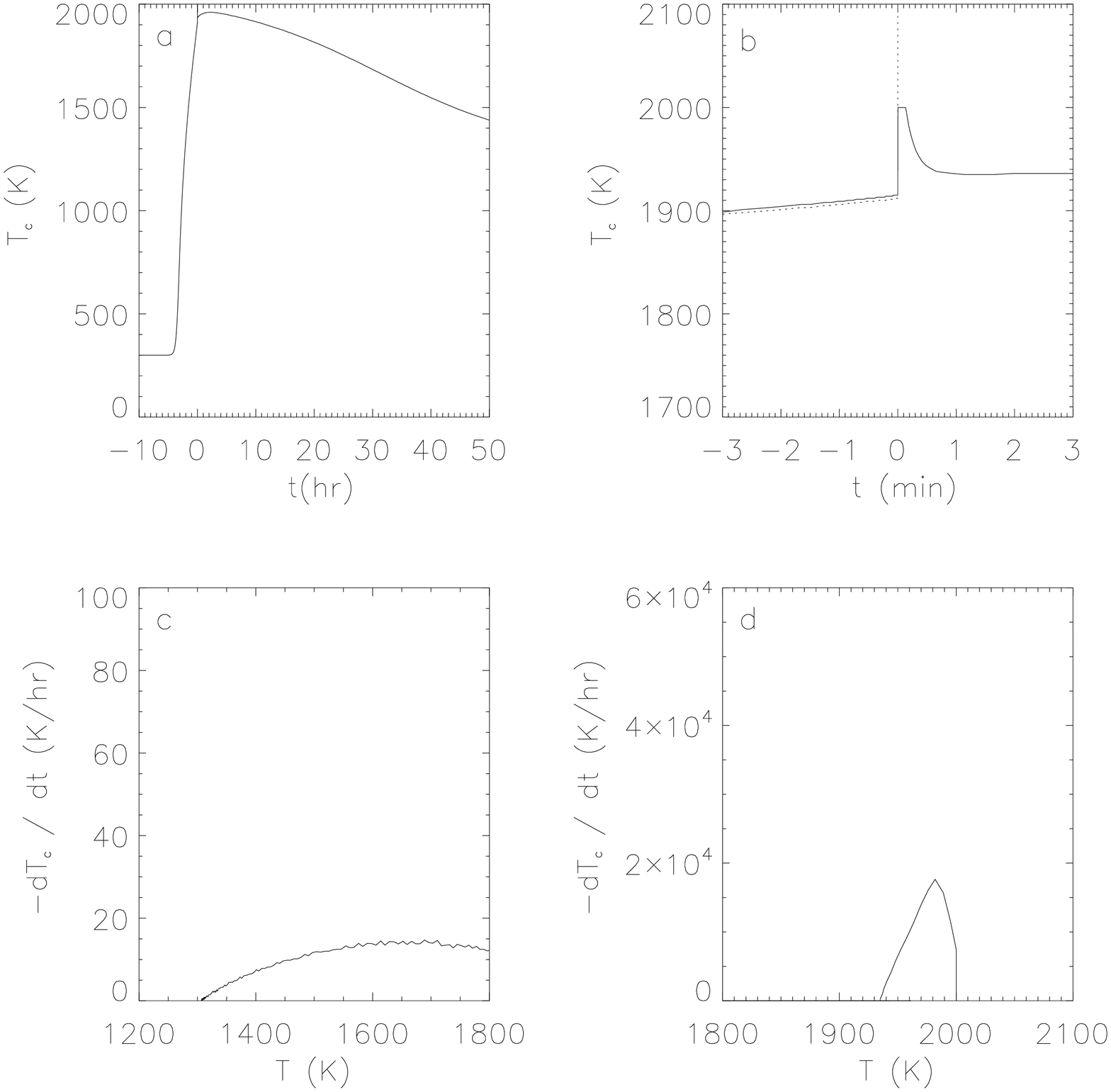}
\caption{Same as Figure~\ref{fig:DC02fig2}.  These results 
were calculated using the original code of DC02, but with $V_s = 8 \, {\rm km} \, {\rm s}^{-1}$ and ${\cal C} = 10$.  (See text for additional shock parameters and details).}
\label{fig:2SD8}
\end{figure}

For comparison to our current results, we also present results of the DC02 code, with a shock speed of $8 \, {\rm km} \, {\rm s}^{-1}$, rather than $7 \, {\rm km} \, {\rm s}^{-1}$, and ${\cal C} = 10$, rather than ${\cal C} = 1$ (Figures~\ref{fig:1SD8} and \ref{fig:2SD8}).
For the $8 \, {\rm km} \, {\rm s}^{-1}$, ${\cal C} = 10$ shock, peak temperatures are seen to be roughly 200 K greater, due to the higher shock speeds.
In fact, the chondrule peak temperatures exceed 2000 K (they are artificially capped at 2000 K in the DC02 model, 
and excess heat is assumed to evaporate the chondrule rather than heat it).
The cooling rates are lower (10 K hr$^{-1}$ vs. 50 K hr$^{-1}$) because in the
${\cal C} = 1$ case dust does not evaporate, but in the $8 \, {\rm km} \, {\rm s}^{-1}$, ${\cal C} = 10$ case the dust evaporates and chondrules take longer to travel several optical depths from the shock front.

\subsection{Updated Code}

\begin{figure}[ht]
\includegraphics[width=14cm]{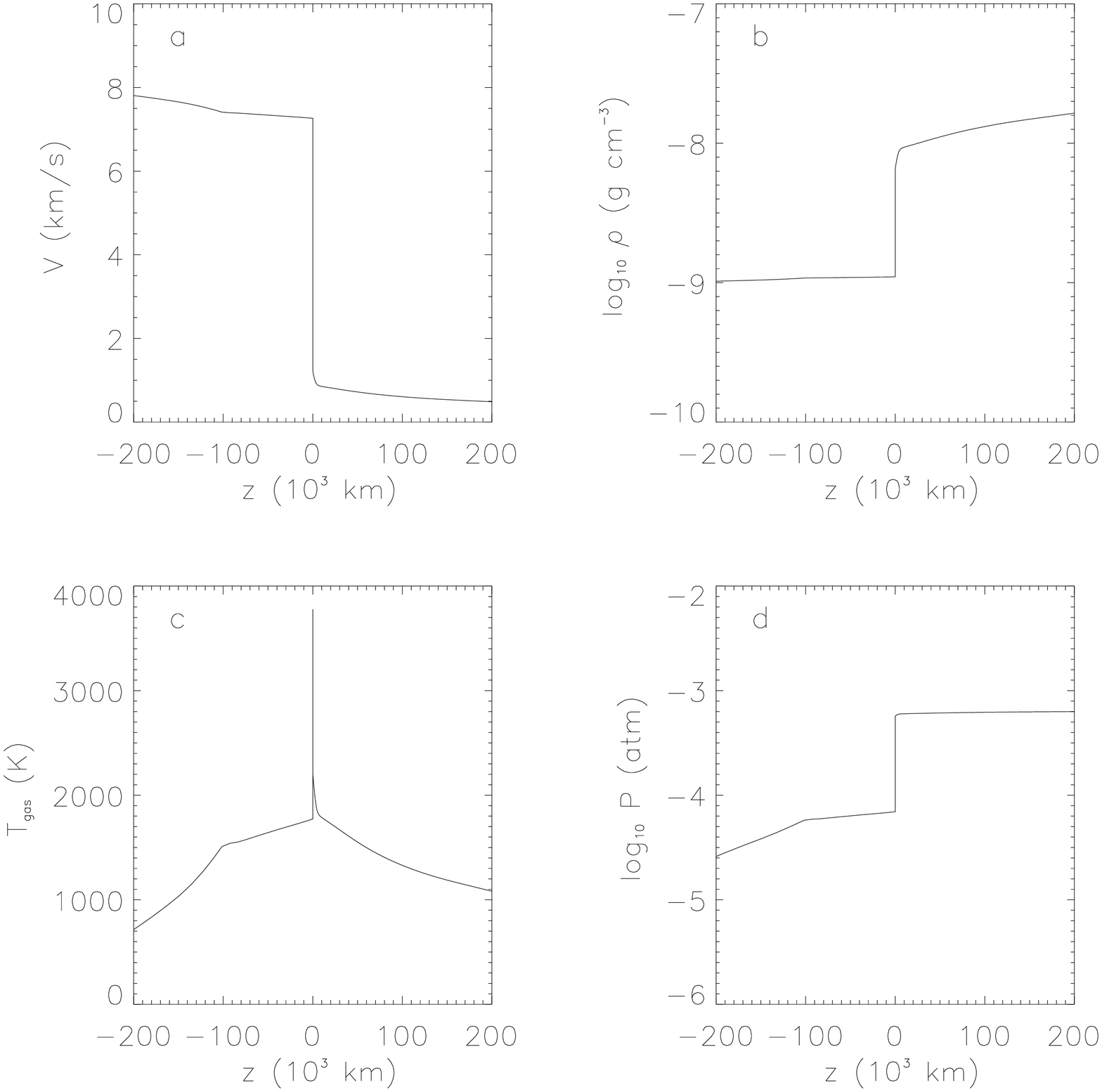}
\caption{Same as Figure~\ref{fig:1SD8}, only these results were calculated using our updated code (with $V_s = 8 \, {\rm km} \, {\rm s}^{-1}$ and ${\cal C} = 10$), and include the effects of higher dust opacity, dust evaporation at 1500 K, and the post-shock boundary condition $T_{\rm post} = T_{\rm pre}$.  (See text for additional shock parameters and details).} 
\label{fig:canonical1}
\end{figure}

\begin{figure}[ht]
\includegraphics[width=14cm]{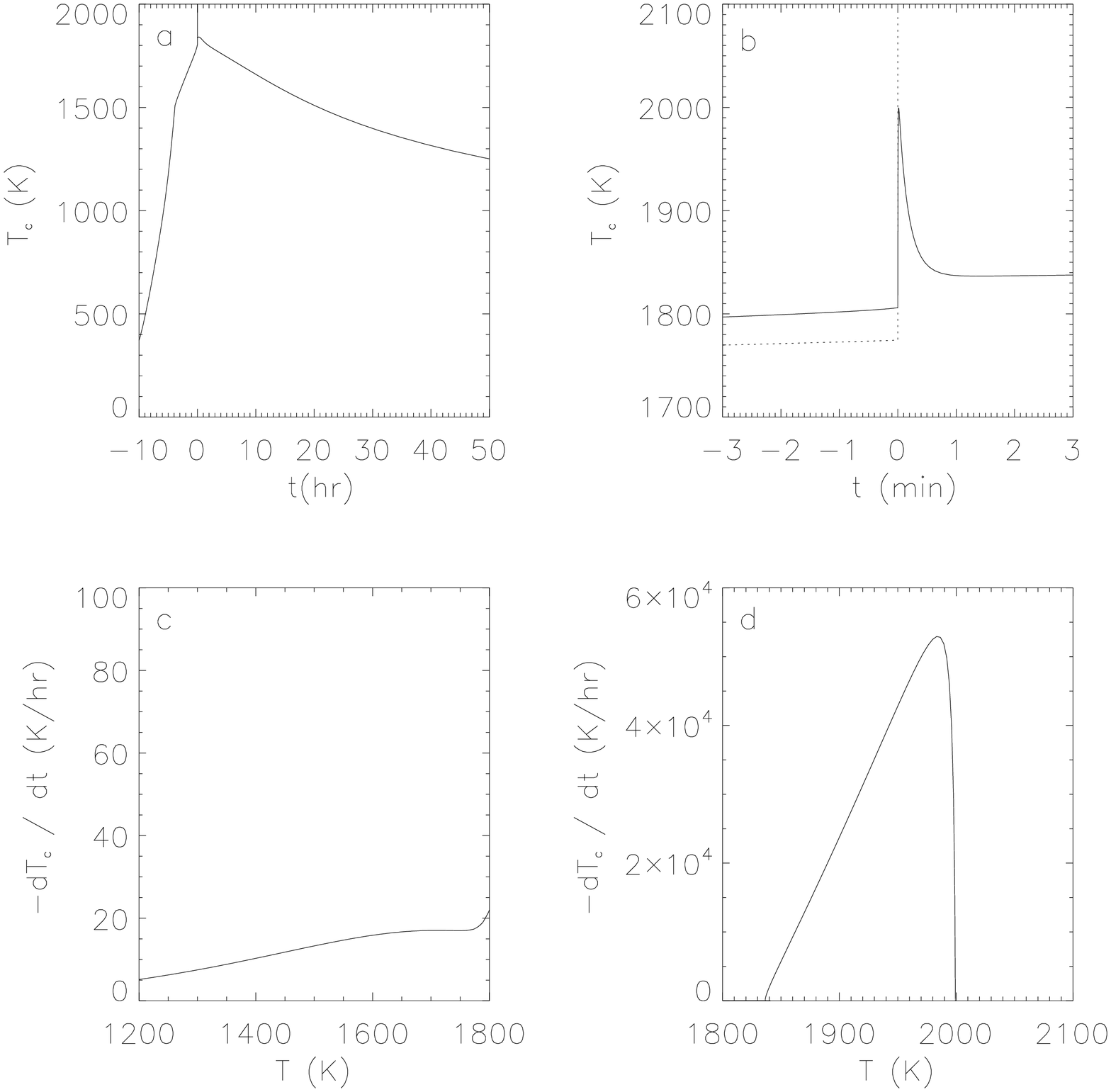}
\caption{Same as Figure~\ref{fig:2SD8}, only these results were calculated using our updated code (with $V_s = 8 \, {\rm km} \, {\rm s}^{-1}$ and ${\cal C} = 10$), and include the effects of higher dust opacity, dust evaporation at 1500 K, and the post-shock boundary condition $T_{\rm post} = T_{\rm pre}$.  (See text for additional shock parameters and details).} 
\label{fig:canonical2}
\end{figure}

We now present the effects of the updates to the code involving dust opacity and evaporation, and the post-shock jump 
conditions. 
Figures~\ref{fig:canonical1} and \ref{fig:canonical2} present the thermal histories of the gas and chondrules including
these effects, but before the inclusion of line cooling.  
Comparing these to the results of DC02 (Figures~\ref{fig:DC02fig1} and~\ref{fig:DC02fig2}), we see that dust evaporates 
much farther into the pre-shock region, at much earlier times.
Because the dust evaporates at 1500 K instead of 2000 K, the optical depths in the pre-shock region are greatly
reduced.  
This, in turn, allows the Marshak wave to diffuse much farther into the pre-shock region (limited mainly by the opacity
of chondrules), allowing temperatures of 1500 K to be reached farther into the pre-shock region. 
The gas and chondrules begin heating up more than 10 hours before they reach the shock front, at an initial rate 
of about $250 \, {\rm K} \, {\rm hr}^{-1}$.
At 3.8 hours before reaching the shock front, the chondrules have already reached 1507 K, and heat up to 1805 K immediately 
before entering the shock.
Their heating rate before hitting the shock, $\approx 100 \, {\rm K} \, {\rm hr}^{-1}$, is lower than the heating rate in
the DC02 model, $\approx 400 \, {\rm K} \, {\rm hr}^{-1}$, due to the lack of significant dust opacity.
Both the dust opacity and evaporation have direct effects on the chondrule thermal histories. 
The post-shock jump condition does not seem to have a large bearing on their thermal histories. 

\begin{figure}[ht]
\includegraphics[width=14cm]{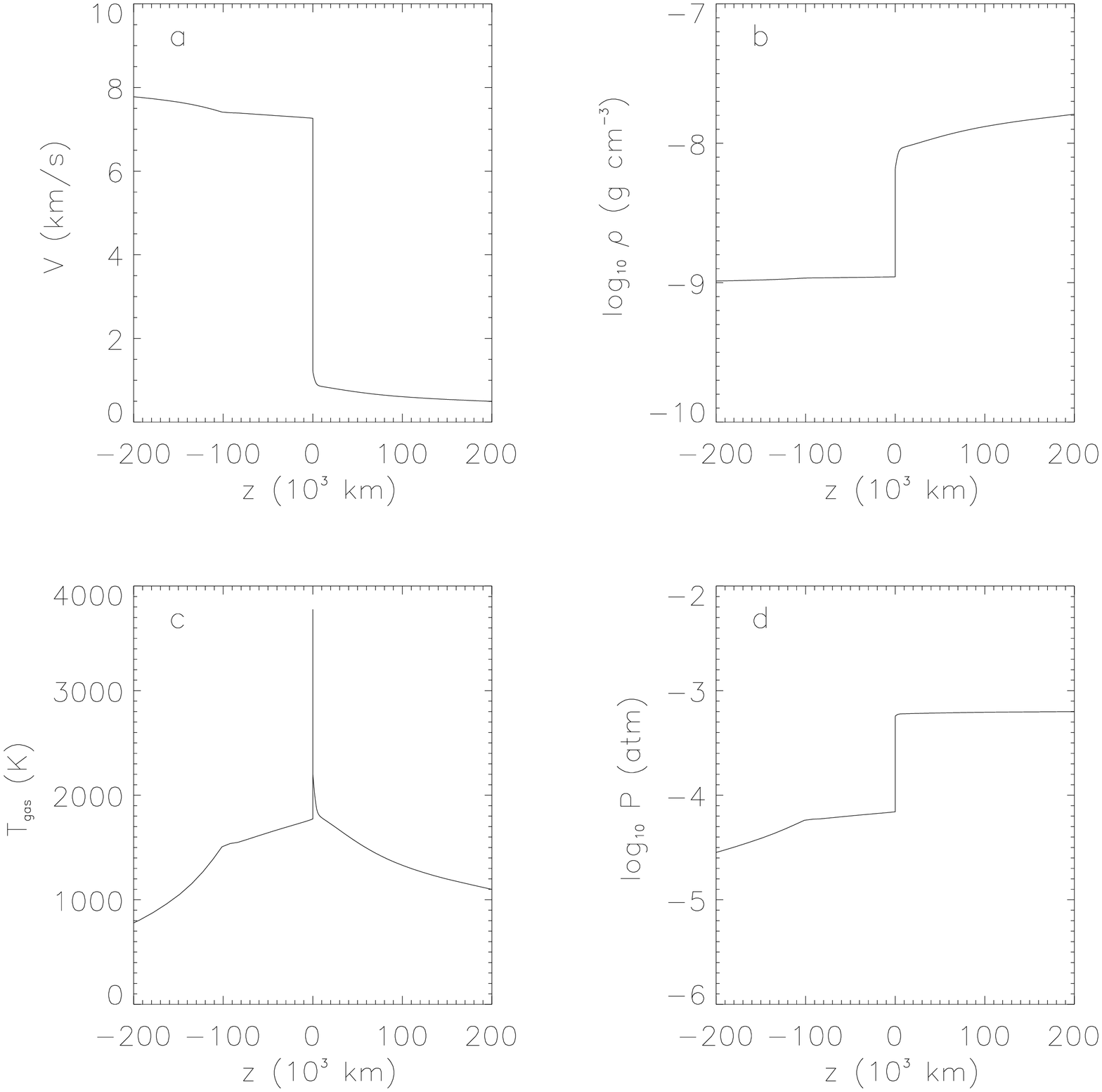}
\caption{Same as Figure~\ref{fig:canonical1}, calculated using our updated code with $V_s = 8 \, {\rm km} \, {\rm s}^{-1}$ and ${\cal C} = 10$.  However, not only do these results include the effects of higher dust opacity, dust evaporation at 1500 K, and the post-shock boundary condition $T_{\rm post} = T_{\rm pre}$, they additionally include line cooling.  (See text for additional shock parameters and details).  This is labeled ``Case 9" in Table~\ref{table:par}.}
\label{fig:fullcool1}
\end{figure}

\begin{figure}[ht]
\includegraphics[width=14cm]{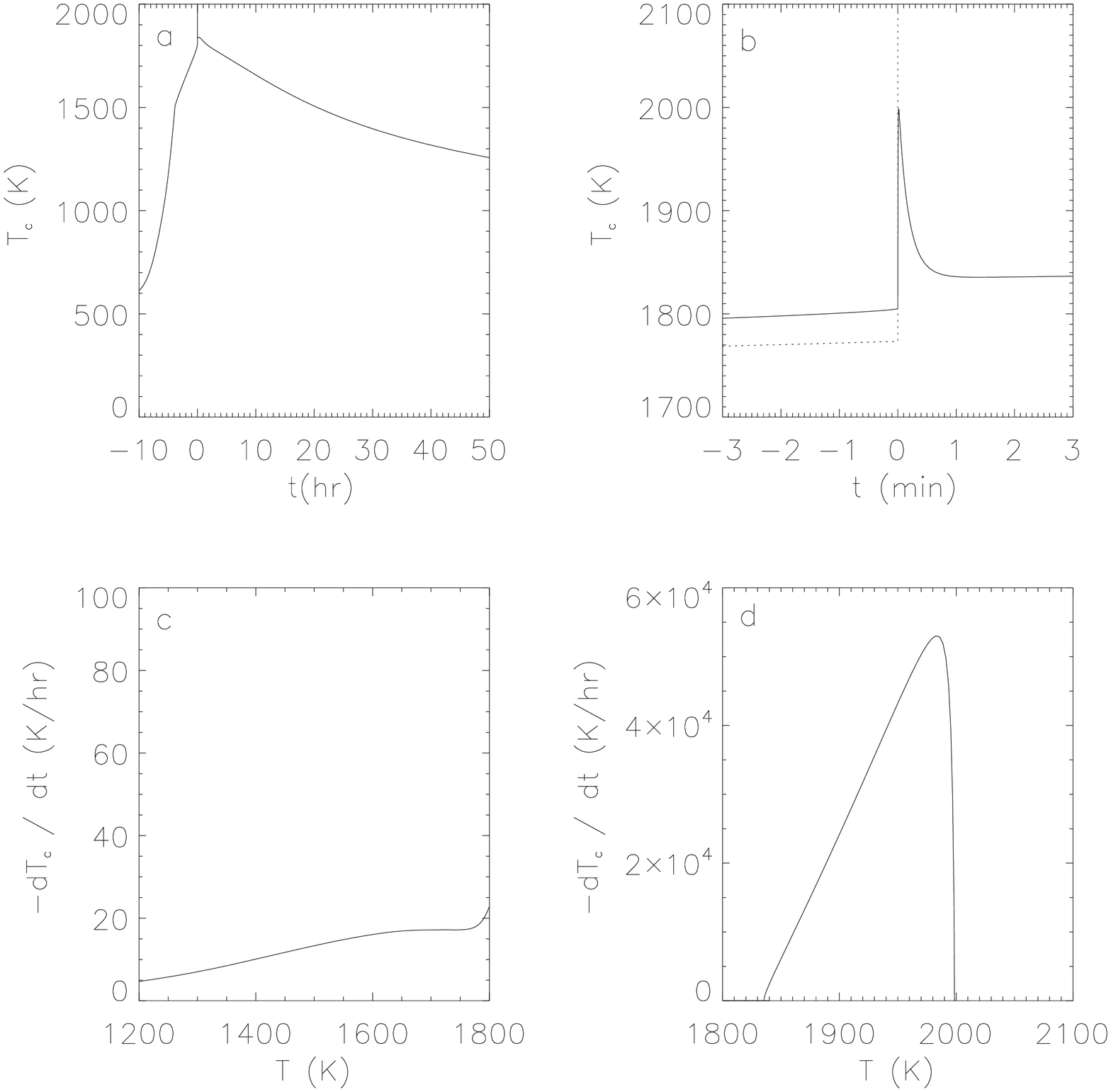}
\caption{Same as Figure~\ref{fig:canonical2}, calculated using our updated code with $V_s = 8 \, {\rm km} \, {\rm s}^{-1}$ and ${\cal C} = 10$.  However, not only do these results include the effects of higher dust opacity, dust evaporation at 1500 K, and the post-shock boundary condition $T_{\rm post} = T_{\rm pre}$, they additionally include line cooling.  (See text for additional shock parameters and details).  This is labeled ``Case 9" in Table~\ref{table:par}.}
\label{fig:fullcool2}
\end{figure}

Next we present the effects of line cooling. 
Based on the toy model of Morris et al.\ (2009), we expected the inclusion of line cooling to result in a noticeable increase 
in cooling rates over our canonical case neglecting line cooling, from \app 10$^2$ K hr$^{-1}$ to \app 10$^4$ K hr$^{-1}$ in
the first $\sim 0.1$ hr after the shock.  
Large dust opacities could in theory reduce the cooling rate, but given that in most of our runs the dust begins to evaporate 
in the pre-shock region, and is completely evaporated at the shock front, dust is unable to arrest the line cooling just
past the shock front. 
The (Planck) opacity in this region (due to chondrules and ultra-refractories) is a mere 0.03 cm$^{2}$ g$^{-1}$, yielding 
photon mean free paths $\approx 5 \times 10^{4} \, {\rm km}$ that are not crossed for tens of hours. 
It is thus surprising that the thermal histories of gas and chondrules in a shock including line emission, depicted in
Figures~\ref{fig:fullcool1} and \ref{fig:fullcool2}, are essentially indistinguishable from Figures~\ref{fig:canonical1} 
and \ref{fig:canonical2}, in which line cooling is neglected.
Line cooling, as it happens, has minimal effects on the thermal evolution of the shock. 

We have conducted a limited parameter study to test the robustness of the conclusion that line cooling does not
significantly affect the thermal histories of gas and chondrules. 
Case 9 is our canonical case, with ${\cal C} = 10$, $V_{\rm s} = 8 \, {\rm km} \, {\rm s}^{-1}$, 
$\rho_{\rm g} = 1 \times 10^{-9} \, {\rm g} \, {\rm cm}^{-3}$, and the water abundance (relative to a 
canonical ratio ${\rm H}_{2}{\rm O} / {\rm H}_{2} = 8 \times 10^{-4}$). 
Our other cases explore variations in these input parameters.
In Table~\ref{table:par} we list these input parameters for each case, and the resultant peak temperature, 
the initial cooling rate following the peak temperature, and the cooling rate through the crystallization 
temperatures (1400 - 1800 K). 

\begin{deluxetable}{cccccccc}
\tablecolumns{8}
\small
\tablewidth{0pt}
\tablecaption{Results of Parameter Study, Including Line Cooling}
\tablehead{
\colhead{Case No.}&
\colhead{${\cal C}$}&
\colhead{V$_s$} & 
\colhead{$\rho_g$ (g cm$^{-3}$)}& 
\colhead{\water\tablenotemark{a}}&
\colhead{ $T_{\rm peak}$ (K)}&
\colhead{Cooling rate \tablenotemark{b}} &
\colhead{Cooling rate \tablenotemark{c}}
 }
\startdata
9& 10& 8& $10^{-9}$& 1& 2000& $>$ 5  \rm x  10$^4$ K/hr& 10-25 K/hr\\
10& 10& 8& $10^{-9}$& 10& 1990& $>$ 5  \rm x  10$^4$ K/hr& 3-30 K/hr\\
11& 10& 8& $10^{-9}$& 0.1& 2000& $>$ 5  \rm x  10$^4$ K/hr& 10-25 K/hr\\
12& 10& 7& $10^{-9}$& 1& 1720& $>$ 2  \rm x  10$^4$ K/hr& 15-45 K/hr \tablenotemark{e}\\
13& 10& 9& $10^{-9}$& 1& 2000\tablenotemark{d}& 1100 K/hr& 8-20 K/hr\\
14& 10& 10& $10^{-9}$& 1& 2000\tablenotemark{d}& 1500 K/hr& 15-21 K/hr\\
15& 10& 8& 3 x $10^{-10}$& 1& 1500& $>$ 7  \rm x  10$^4$ K/hr& 7 x 10$^4$ K/hr\\
16& 30& 8& $10^{-9}$& 1& 1990& $>$ 5  \rm x  10$^4$ K/hr& 22-65 K/hr\\
17& 50& 8& $10^{-9}$& 1& 2000\tablenotemark{d}& $>$ 10$^4$ K/hr& 22-80 K/hr\\
\enddata
\tablenotetext{a}{Water abundance with respect to our assumed water-to-gas ratio, 8 x 10$^{-4}$.}
\tablenotetext{b}{Cooling rates at $T_{\rm peak}$}
\tablenotetext{c}{Cooling rates through 1400-1800 K, the crystallization temperature range of chondrules.}
\tablenotetext{d}{Artificial peak temperature due to evaporation.}
\tablenotetext{e}{Cooling rates through 1400-1599 K; rates are \app 2 x 10$^4$ from 1600-1800 K.}
\label{table:par}
\end{deluxetable}

As is evident from Table~\ref{table:par}, the effect of increased shock speed is to increase $T_{\rm peak}$.
As $V_{\rm s}$ increases from 7 to 8 to 9 to $10 \, {\rm km} \, {\rm s}^{-1}$ (cases 12, 9, 13 and 14), 
$T_{\rm peak}$ increases from 1720 K to 2000 K, to temperatures exceeding 2000 K.
The chondrule temperatures were not allowed to exceed 2000 K, and instead begin to undergo evaporation
(see DC02).
Because of this, the chondrules do not drop below 2000 K until after the gas's rapid cooling phase is over,
explaining the lower cooling rates from the ``peak" in cases 13 and 14.

The effect of lowering the gas density is equally clear.  
When the gas density is lowered by a factor of 3 (case 15), the frictional drag on chondrules is reduced,
and the peak temperature reaches only 1500 K. 
As such, the rapid cooling from the peak takes place in the chondrule's crystallization temperature range.

The effect of increasing the chondrule concenration is seen in cases 9, 16 and 17, where ${\cal C}$ 
increases from 10 to 30 to 50. 
The differences in $T_{\rm peak}$ between these cases are not great ($<$ 10 K).
We expected to see a rise of \app 50 K between ${\cal C}$ = 10 and ${\cal C}$ = 50, based on the observations and predictions of DC02.  They observed a distinct increase in $T_{\rm peak}$ with increased chondrule concentration, but we don't observe that behavior in our results.  
This is, in large part, because the temperatures are capped at 2000 K.  So although we don't see quite the pronounced trend that DC02 observed and predicted, we can't rule that out.  
  
The most surprising result pertains to the variation in water density. 
In cases 11, 9 and 10 this density is increased from 10\% of the canonical value, to the canonical value,
to 10 times the canonical value of ${\rm H}_{2}{\rm O} / {\rm H}_{2} = 8 \times 10^{-4}$. 
Yet even when the water abundance is increased by a factor of 10, the cooling rates of chondrules (over 
their crystallization temperature range) are increased only by $\approx 10 \, {\rm K} \, {\rm hr}^{-1}$.
This increased cooling rate falls considerably short of the cooling rates $\sim 10^{4} \, {\rm K} \, {\rm hr}^{-1}$
predicted by the toy model of Morris et al.\ (2009), indicating that a physical effect not included in the toy
model is preventing the expected increase in the cooling rate due to line emission.  
We investigate this issue in what follows.

\section{Discussion}

In this section, we seek a physical explanation for the lack of significant cooling due to emission of line photons.
The toy model of Morris et al.\ (2009) included line cooling but did not consider hydrogen dissociation and 
recombinations, and did not consider changes in the gas velocity or density.
This model had predicted that gas more than a few hundred kilometers past the shock front would be surrounded in
all directions by sufficient column densities of warm gas to prevent the escape of line photons, and line
cooling would effectively shut off.  
But travelling this distance would take roughly 0.1 hr, during which time gas would be cooling at rates 
$\sim 10^3 - 10^{4} \, {\rm K} \, {\rm hr}^{-1}$, dropping several hundred K in temeprature. 
Chondrules are imperfectly thermally coupled to the gas during this stage, but might be expected to 
drop several hundred K in temperature as well. 
The runs presented here did not exhibit this behavior, though, as chondrule temperatures differed by only
a few K due to the inclusion of line cooling.
Some physical mechanism not included in the toy model must act on chondrules heated in shocks. 

Dissociation and recombinations of hydrogen molecules are already known to have a major bearing on 
the gas temperature in chondrule-forming shocks, especially in the first seconds after passing through
the shock front. 
Each dissociation of an ${\rm H}_{2}$ molecule consumes $4.48 \, {\rm eV}$ of energy. 
This is about 10 times the thermal energy of the molecule, so dissociation of just a few percent of 
the ${\rm H}_{2}$ molecules can significantly change the temperature of the gas. 
Indeed, in our canonical shock (without line cooling), after the adiabatic jump across the shock front
raises the gas temperature to 3800 K, dissociation of hydrogen molecules occurs,
cooling the gas, so that quasi chemical equilibrium is reached, where \app 15 \% of the hydrogen molecules are dissociated and the gas is now at 2200 K.
This cooling takes place in approximately 17 seconds, at a rate
$\sim 10^{5} \, {\rm K} \, {\rm hr}^{-1}$. 
Since hydrogen dissociation also buffers the temperature rise at the shock front,
it can be expected that as the gas cools, recombinations of hydrogen molecules, which will release
energy, will buffer the cooling as well. 

We can estimate the buffering of cooling rates while hydrogen molecules are recombining. 
We suppose a purely hydrogen gas with a number density of hydrogen nuclei (atomic plus
molecular) $n_{\rm H,tot} = 2 n_{\rm H2} + n_{\rm H}$.
We define the atomic fraction $f = n_{\rm H} / n_{\rm H,tot}$. 
Following equation 6.1 of Spitzer (1978), we now compute the thermal evolution of the gas as 
\begin{equation}
n_{\rm H} \, \frac{d}{dt} \left( \frac{3}{2} k T \right) -k T \, \frac{d n_{\rm H}}{d t}
= \left( \Gamma - \Lambda \right)_{\rm rad}
- n_{\rm H,tot} \, \left( \frac{d f}{dt} \right) \, \frac{\epsilon}{2} \nonumber
- n_{\rm H2} \, \frac{d}{dt} \left( \frac{5}{2} k T \right) + k T \, \frac{d n_{\rm H2}}{d t}, 
\end{equation}  
where $\epsilon = 4.48 \, {\rm eV}$ is the dissociation energy of ${\rm H}_{2}$,
and $(\Gamma-\Lambda)_{\rm rad}$ is the net radiative heating rate of the gas. 
By assuming constant density in the post-shock region ($dn/dt = 0$), 
it is straightforward to show that 
\begin{equation}
\frac{dT}{dt} =
 \left[ \frac{5+f}{2} + \left( T \frac{\partial{f}}{\partial{T}} \right)_n \,\left(\frac{\epsilon}{kT}-1 \right)  \right]^{-1} 
 \, \left[  \frac{2 \left( \Gamma - \Lambda \right)_{\rm rad}}{n_{\rm H,tot}\;k} \right].
\end{equation}
Compared to the case without dissociation / recombination, in which $f \equiv 0$, the cooling 
rate is reduced by a ``buffering factor" $B$, where   
\begin{equation}
B = \frac{ 1 }{ 1 + f } \, \left[ 1 + \frac{1}{5}f + \frac{2}{5} \left( T \frac{\partial{f}}{\partial{T}} \right)_n \left(\frac{\epsilon}{kT}-1 \right)  \right].
\label{eq:buffering} 
\end{equation}
The buffering factor can usually be approximated as 
\begin{equation}
B \sim 0.4 \, \left( \frac{\epsilon}{k T} \right) \, \left( T \frac{\partial{f}}{\partial{T}} \right).
\end{equation}
We can better estimate the buffering of cooling rates due to recombinaiton by determining the atomic fraction $f$ as a function of temperature.
Utilizing the kinetic rates of ${\rm H}_{2}$ dissociation and recombination listed in DC02, and 
assuming an overall density $n_{\rm H,tot} = 2.57 \times  10^{15} \, {\rm cm}^{-3}$ 
(equivalent to $\rho_{\rm g} = 6 \times 10^{-9} \, {\rm g} \, {\rm cm}^{-3}$),
we have calculated the equilibrium atomic fraction $f$ as a function of temperature.
(It is worth noting that chemical equilibrium is established within a matter of seconds at these
densities.)
Both $f(T)$ and its derivative $T (\partial{f}/\partial{T})$ are displayed in Figure 10.
These are used as inputs to the buffering factor of Equation~\ref{eq:buffering}, which is displayed
as a function of temperature in Figure 11. 
It is readily seen that at constant density (approximately true in the post-shock region), the heating due to hydrogen molecule recombinations
reduces the cooling rates by factors \app 2-10 below what they would otherwise be. 

\begin{figure}[ht]
\includegraphics[width=14cm]{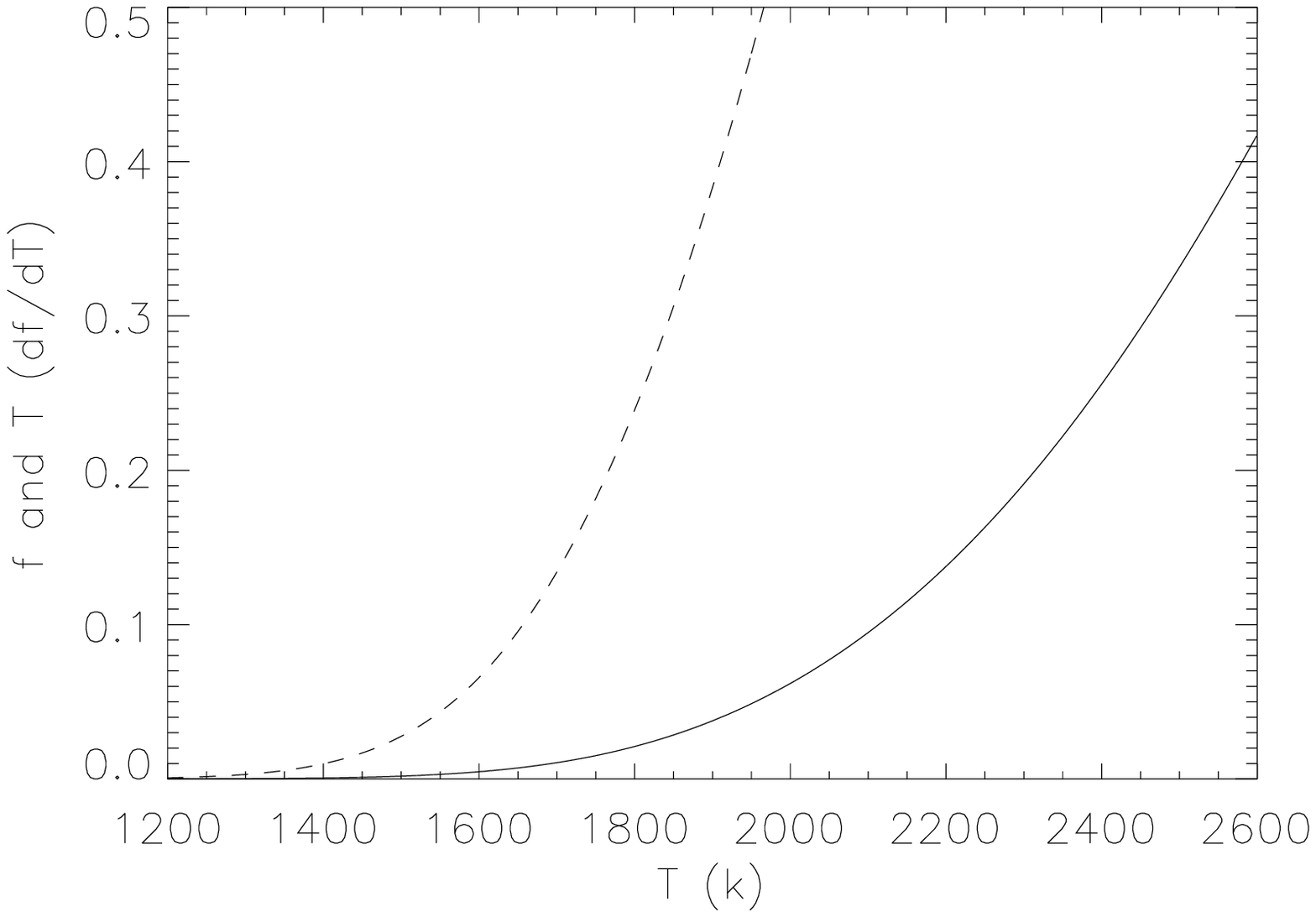}
\caption{The equilibrium atomic fraction $f$ (solid line) and its derivative $T(\partial{f}/\partial{T})$ (dashed line) as a function of 
temperature $T$.}
\label{fig:df}
\end{figure}

\begin{figure}[ht]
\includegraphics[width=14cm]{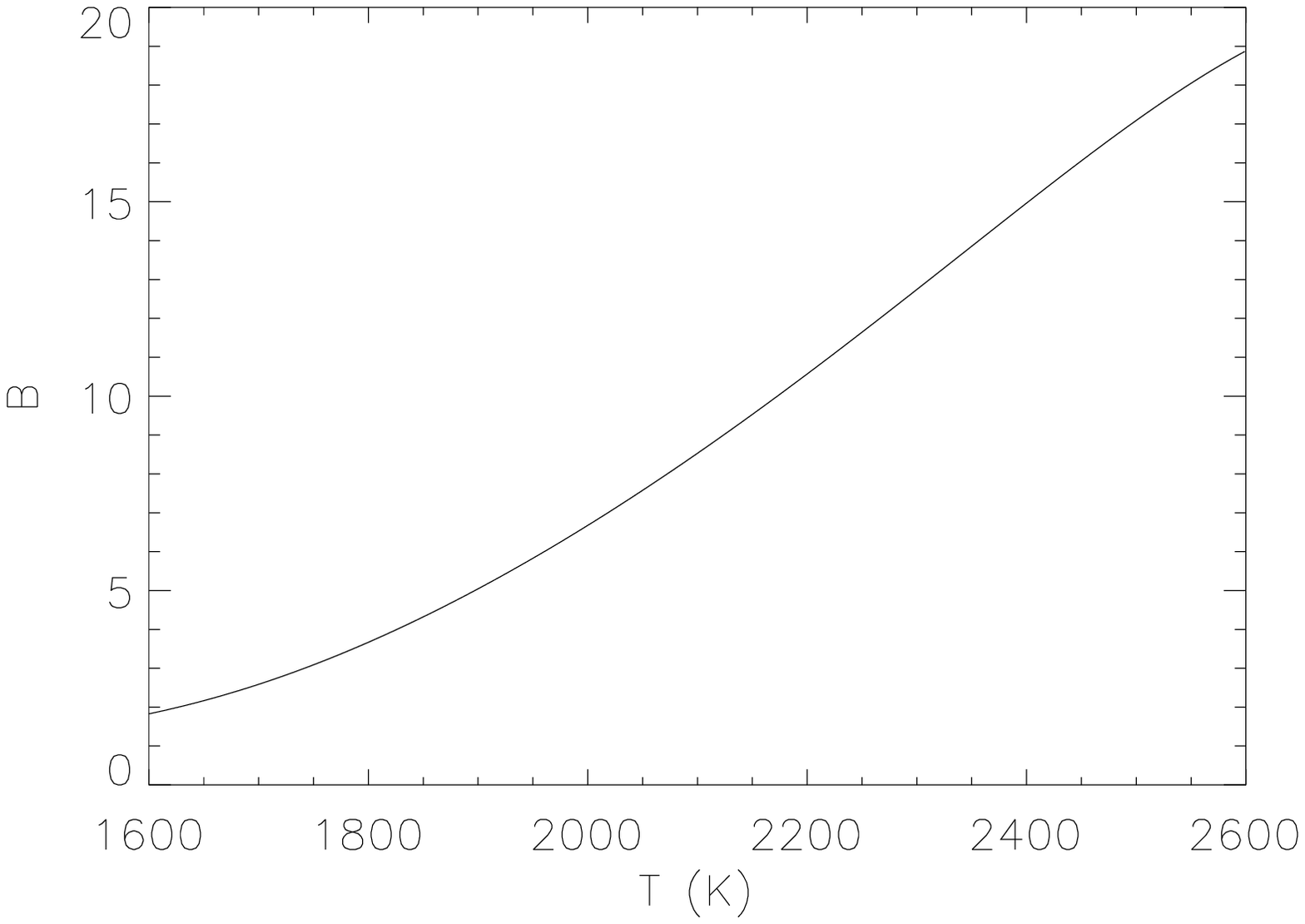}
\caption{The buffering factor $B$ by which cooling rates are reduced due to thermal buffering by ${\rm H}_{2}$ 
recombinations.  In the relevant temperature range, cooling rates are reduced by a factor of 2-10.}
\label{fig:net}
\end{figure}

Buffering by recombination of ${\rm H}_{2}$ molecules, in concert with other
factors, conspires to reduce the efficacy of line cooling.
If one were to estimate the cooling rate of the gas due to optically thin line emission, 
one might take the cooling rate per volume to be $n_{\rm H_2O} {\cal L}$, where  
$n_{\rm H_2O}$ is the number density of water molecules, and 
${\cal L}$ = $8.3 \times 10^{-12} \, {\rm erg} \, {\rm s}^{-1}$ 
at 2250 K (Morris et al.\ 2009). 
The thermal energy of the gas is $(5/2) n_{\rm H2} \, k T + (3/2) n_{\rm He} \, k T$.
Assuming ratios $n_{\rm He} / n_{\rm H2} = 0.2$ and $n_{\rm H_2O} / n_{\rm H2} = 8 \times 10^{-4}$, 
one would estimate a cooling rate $dT / dt \approx 6.1 \times 10^{4} \, {\rm K} \, {\rm hr}^{-1}$
at 2250 K.
One effect that reduces the cooling rate below these values is the column density of water.
The first few seconds past the shock are dominated by the temperature drop associated with
${\rm H}_{2}$ dissociation and are characterized by a decoupling of chondrule and gas
temperatures; at any rate, the chondrule temperature does not record conditions in the
first few seconds past the shock.
A mere 10 seconds (and about 10 km) past the shock front, though, (during which line cooling 
could cool the gas by at most $\approx 150 \, {\rm K}$), the column density of water between gas and the
shock front is $N_{\rm H_2O} \approx 10^{18} \, {\rm cm}^{-2}$, already leaving the optically
thin regime of line emission.
This alone will reduce the cooling rate at 2250 K to $4.3 \times 10^{4} \, {\rm K} \, {\rm hr}^{-1}$.
A second effect is backwarming.
Although the hot, post-shock gas ($\sim$ 2200 K) is emitting \wwater line emission, \wwater molecules in this region also absorb line radiation from surrounding ``warm" regions ($\sim$ 1750 K, both in the pre-shock and post-shock regions).  So the net cooling is reduced.
As we discuss below, this hot slab of gas will also heat gas further downstream. 
A third effect reducing the cooling rate is the buffering of the cooling rate by hydrogen
recombinations.  
At 2250 K (just 10 s past the shock front), these reduce the cooling rate by a factor $B \approx 8.0$, to only 
$3800 \, {\rm K} \, {\rm hr}^{-1}$.

As the gas moves farther from the shock front and continues to cool, all the factors conspire to
reduce the cooling rate from line emission: the temperatures drop, lowering the rate of emission;
the column density of water to the shock front increases; the relative effects of backwarming
increase; and the buffering from hydrogen recombinations remains effective. 
For example, as the gas moves to a distance 3000 km past the shock front after 3000 s,
the cooling rate of the gas drops considerably as the column density of water 
($= 3 \times 10^{20} \, {\rm cm}^{-2}$) becomes optically thick, to only $160 \, {\rm K} \, {\rm hr}^{-1}$.
This implies that the effect of line cooling should be to make the gas drop in temperature during the
first hour, by an extra several hundred K above what it would otherwise drop. 
Cooling rates $\sim 10^{2} \, {\rm K} \, {\rm hr}^{-1}$ are much slower than one would predict 
in the absence of a finite column density of water and the buffering due to hydrogen recombinations.
On the other hand, our complete calculation suggests the effect of line cooling is even smaller than
this, leading only to an additional $\approx 3 \, {\rm K} \, {\rm hr}^{-1}$. 
This indicates that the reasons for the slow cooling lie in additional factors still not considered. 

The gas cooling rate, we find, is extremely insensitive to the effects of line cooling.
We attribute this to the complicated nature of ``backwarming" in the shocked flow.
At every point, even immediately past the shock front, \wwater molecules emit into a background 
only slightly cooler than the local temperature.  
Since there is nowhere much cooler to radiate, the net cooling is much reduced.
Line photons tend to be absorbed within about 100 km of the place where they are emitted. 
Based on the temperature gradient far downstream from the shock, the temperature change over
100 km is $< 0.3 \, {\rm K}$.
Because of backwarming, water molecules will on average absorb line photons from the slightly 
warmer upstream gas, and on average emit line photons to the slightly cooler downstream gas, 
but the net effect is very small.
Just past the shock front, the temperature gradients are larger, but the heating is more intense.
The shocking of the gas produces a hot slab of gas with a thickness of hundreds of kilometers and a 
temperature well in excess of 2000 K, which emits line photons into the post-shock gas.
We attribute the extremely reduced gas cooling rates we observe in our simulations 
(only $\approx 3 \, {\rm K} \, {\rm hr}^{-1}$ above what they would be otherwise, at 100 s past the 
shock front), to the fact that line emission only cools the gas if line photons can escape into a 
cooler medium, and in the shocks considered here, there are no significantly cooler regions for the
line photons to be lost to.
Significantly, this result could not be discovered without the approach we have adopted here. 

The effect of line emission on chondrule temperatures is even more limited.
After the drag heating stage of chondrules ends, at about 30 seconds, the temperatures of chondrules are primarily 
set by absorption and emission of radiation.
The rate at which chondrules absorb or emit radiation is, per area, $\approx 0.8 \sigma T_{\rm c}^{4}$
$\approx 5.3 \times 10^8 \, {\rm erg} \, {\rm cm}^{-2} \, {\rm s}^{-1}$ at $T_{\rm c} = 1850 \, {\rm K}$. 
This is to be compared to the rate at which they thermally exchange energy with the gas,
$\approx (5/2) n_{\rm H2} \, k (T_{\rm g} - T_{\rm c}) \, (v_{\rm th} / 4)$,
where $v_{\rm th} = (8 k T_{\rm g} / \pi m_{\rm H2})^{1/2}$ is the thermal speed of the molecules.
At a gas temperature $T_{\rm g} = 2200 \, {\rm K}$, $v_{\rm th} = 4.8 \, {\rm km} \, {\rm s}^{-1}$.
Assuming a post-shock density $6 \times 10^{-9} \, {\rm g} \, {\rm cm}^{-3}$ and a purely ${\rm H}_{2}$
gas, the thermal exchange rate is only $2.6 \times 10^{7} \, {\rm erg} \, {\rm s}^{-1}$. 
That is to say, the absorption of heat energy from the gas only amounts to about 5\% of a chondrule's 
overall energy budget; 95\% of its heating comes from absorption of the radiation from other chondrules.
Line emission can lower the temperature of the gas by a few K in the first few hours past the shock.
Say the temperature drop is 6 K: then the loss of heating from the gas amounts to about 3\% of the
chondrule's total heating, reducing the chondrule's temperature by $<$ 1 K over the course of perhaps
3 hours.  
This is consistent with the difference in chondrule cooling rates $\approx 0.3 \, {\rm K} \, {\rm hr}^{-1}$ 
we observed in the temperature range 1800 - 1900 K. 
At later times, as the gas and chondrules reach thermal equilibrium and line cooling shuts off more, the
effect of line cooling on the cooling rates is lessened even further. 

Line cooling is seen to have minimal effect on the cooling rates of chondrules melted in chondrule-forming
shocks. 
Even when the water abundance is increased by a factor of 10, chondrule cooling rates increase by only
$\approx 30 \, {\rm K} \, {\rm hr}^{-1}$ in our runs (and an even smaller increase, 
$\sim 3 \, {\rm K} \, {\rm hr}^{-1}$, for chondrules).
We attribute this primarily to two effects: the backwarming due to heat from surrounding regions radiating
into the post-shock region (which reduces line cooling), and especially the buffering effects of the heat 
released by the recombination of hydrogen molecules.
Had these effects not been considered, the gas temperature in the first few hours past the shock would have 
dropped by $\sim 10^{2} - 10^{3}$ K more than we observed it to.
The gas then would be significantly cooler than the chondrules, and would have produced a strong cooling 
effect on the chondrules.
Especially after a few hours, when the chondrules and gas are in thermal equilibrium, both components 
would have cooled to the background temperature, implying a cooling rate $\sim 10^{3} \, {\rm K} \, {\rm hr}^{-1}$.
The buffering effects of ${\rm H}_{2}$ recombination prevent this scenario, reducing the effects of 
line emission, resulting in chondrule thermal histories almost identical to those without line cooling.

Previously, Scott et al.\ (1996) had suggested that dissociations of hydrogen molecules would limit
the peak temperatures of gas and chondrules as thermal energy was used up breaking the bonds of 
${\rm H}_{2}$ molecules.
Our modeling confirms that, but also demonstrates the comparable magnitude of the recombinations
of ${\rm H}_{2}$ molecules.  
As atomic hydrogen recombines, the chemical energy released is converted into thermal energy, which
buffers the cooling rate of the gas. 
Our modeling suggests the cooling rates are reduced by up to one order of magnitude by this effect.
We note that Iida et al.\ (2001) and Miura \& Nakamoto (2006) considered hydrogen dissociations and recombinations 
in the energetics of the gas.  
Significantly, though, they did not calculate the heating and cooling rates by considering a single
equation for the rate ${\rm H} + {\rm H} \rightarrow {\rm H}_{2}$ or its inverse reaction. 
For dissociation, they used equation B6 of Iida et al.\ (2001), which considers dissociation of 
hydrogen molecules by collisions with ${\rm H}_{2}$, H, and electrons.
For recombinations, they used equation B1 of Iida et al.\ (2001), which considered recombinations
by three-body reactions (${\rm H} + {\rm H} + {\rm H}$) and ionized species, but not three-body reactions
involving hydrogen molecules (${\rm H}_{2} + {\rm H} + {\rm H}$).  
In most astrophysical settings in which hydrogen recombinations are considered, ${\rm H}_{2}$ molecules are
relatively rare, but in the nebular setting of chondrule-forming shocks, they are the dominant species.
In any case, the reaction networks implicitly assumed by the formulas employed by Iida et al.\ (2001)
do not match, and potentially could lead to spurious results.
Calculating the atomic fraction of hydrogen accurately is important because, as our calculations show, hydrogen recombinations can reduce the cooling rates by an order of magnitude. 

We find that the combined effects of backwarming and hydrogen recombinations mean that line cooling
can increase the cooling rates of the gas, but only by about $3 \, {\rm K} \, {\rm hr}^{-1}$ for 
standard parameters.
The effect on the cooling rates of chondrules is even lower, amounting to only 
$\approx 0.3 \, {\rm K} \, {\rm hr}^{-1}$, typically. 
We find that the nebular shock model remains largely consistent with the thermal histories of chondrules.  
In the shock model, chondrule precursors are initially at the low temperature necessary for the retention 
of primary sulfur, are heated rapidly to peak temperatures consistent with observational evidence, experience 
rapid initial cooling and then slower cooling through their crystallization temperatures, at rates consistent 
with experimental constraints on their thermal histories. 

One potential weakness of the shock model for chondrule formation, however, is that the shock wave model 
remains insconsistent with chondrule constraints regarding isotopic fractionation in the pre-shock region.  
Recall that there is no indication of the isotopic fractionation in chondrules that would arise from the free 
evaporation of alkalis, which constrains the time spent at high temperature before melting (Tachibana et al.\ 2004) 
to minutes or less (Tachibana \& Huss 2005).  
We see no way around extended pre-heating of chondrules in a 1-D shock, because in 1-D a Marshak wave must propagate 
into the pre-shock region. 
The Marshak wave must propagate a number of (infrared) optical depths into the pre-shock region so that the
radiative diffusion time is comparable to the travel time; this implies optical depths $\sim 10^{2} - 10^{3}$.
A higher opacity in the pre-shock region could reduce the time particles spend at high temperatures; but 
the only way to achieve such high opacities is to have dust grains or an expectionally high density of chondrules.
Dust grains are virtually certain to evaporate completely in the pre-shock region in a chondrule-forming shock.
On the other hand, the density of chondrules that would significantly increase the opacity are about 3 orders of
magnitude more dense than the average density of solids in the solar nebula.
An alternative solution is that the assumption of 1-D may be wrong.
Planetesimal bow shocks would not be large-scale structures that would generate a Marshak wave that penetrates 
deeply into pre-shock region. 
Further experimental work clarifying what the lack of isotopic fractionation means, as well as modeling of 2-D 
shocks, are warranted.
Such studies may discriminate between planetesimal bow shocks and gravitational instabilities as the sources 
of solar nebula shocks. 

One other constraint that is difficult to interpret in the context of the shock model is the recent discovery 
of primary Na in olivine phenocrysts in chondrules.  
Maintaining a high partial pressure of Na vapor in the gas, as the chondrules cooled, is the only way to stabilize 
the chondrule melt against the loss of this volatile.  
Although our model predicts evaporation of some fraction of the chondrule material ($\sim 10\%$), for typical nebular
densities of chondrules, the chondrule vapor density is insufficient to stabilize the chondrule melt. 
Alexander et al.\ (2008) invoke high concentrations of chondrules, $> 10^4 - 10^5$ times nebular densities, 
even though they acknowledge that the chondrule densities necessary are unrealistically high, based on our current 
understanding of the solar nebula.
For example, settling to the midplane limits chondrule concentrations to $\sim 10^2$, because higher concentrations 
induce shear instabilities and mixing that prevent further settling (Sekiya \& Nakagawa 1988).
Models of turbulent concentration (Cuzzi et al.\ 2008) do predict that chondrules should collect in clumps, some of
which have concentrations $\sim 10^4$, although only on scales $< 10^3$ km.
Such small regions are expected to be optically thin and cool too rapidly to be consistent with chondrule textures,
but it must be admitted that the transfer of radiation in these regions can only be modeled in 2-D.
In the context of the 1-D shock model, there is no way to explain the primary Na in olivine phenocrysts in
chondrules. 

As 2-D shock models are developed, however, we envision a possible explanation for the high partial pressure of Na in 
chondrule-forming regions.
Building on the models of Cuzzi et al.\ (2008), we assume there are regions of the nebula $\sim 10^3 \, {\rm km}$
in extent, in which ${\cal C} \sim 10^4$, which are surrounded by a ``penumbra" of lower concentrations (${\cal C} \sim 10-100$) 
on much larger scales, $\sim 10^4 - 10^5 \, {\rm km}$. 
This extended region is overrun by a shock of a given speed, say $8 \, {\rm km} \, {\rm s}^{-1}$. 
As predicted by DC02, and as shown by our results, higher ${\cal C}$ regions would also result in higher peak temperatures 
and faster cooling rates (see Table 1).
For sufficiently high ${\cal C}$, DC02 have shown the concentrations in the heart of the clump are sufficient 
to completely evaporate all chondrules in the densest part, while only melting chondrules in the penumbral regions.
In the hours during which melted chondrules are cooling and recrystallizing, they can come into contact with
the chondrule vapor from the densest region.
This could occur if the chondrule vapor can move outward, for example by pressure-driven expansion of the
chondrule vapor, which can carry the vapor thousands of kilometers in a few hours. 
An alternative possibility is that chondrules are {\it focused} into the clumps. 
In clumps, chondrules are a significant fraction of the gas mass, increasing the overall density of material in 
this region, so the shock ends up propagating more slowly through the clump than the surrounding gas.
The trajectories of chondrules entering the shock are refracted, and chondrules are focused into the clump, 
{\it after} the clump has experienced its peak heating. 
Therefore, some fraction of chondrules should experience otherwise normal 
thermal histories indicative of moderate chondrule concentrations, but in
the presence of very high pressures of chondrule vapor that can only arise
from regions of higher chondrule concentration.

This scenario may resolve the quandary of Alexander et al.\ (2008).
Because chondrules in the clumps totally evaporate, the requisite Na vapor pressure can be met by evaporating 
a solids density in the post-shock region of only $1 \, {\rm g} \, {\rm m}^{-3}$.  
This equates to about $0.1 \, {\rm g} \, {\rm m}^{-3}$ before the shock. 
For $300 \, \mu{\rm m}$-sized chondrules, this implies a chondrule density of $ \approx 300 \, {\rm m}^{-3}$
and ${\cal C} \sim 10^{4}$ in the pre-shock gas.
This is near the upper limit of chondrule concentrations thought attainable by turbulent concentration, but still 
achievable, and is consistent with the scenario envisioned here.
A more quantitative analysis to test this hypothesis requires a shock code that can handle motions parallel to 
the shock front as well as radiative transfer in 2-D cylindrical geometry.

\section{Conclusion}

In this paper, we have presented the results of our updated shock model for chondrule foramation. 
We have described our updates to the hydrodynamic shock code of DC02, correcting the problems with 
previous shock codes identified by Desch et al.\ (2005); namely, the appropriate boundary condition 
for the input radiation and the proper method for calculation of the opacity of solids, and the 
inclusion of a complete treatment of molecular line cooling due to water.

We have found that using the appropriate boundary condition for the input radiation, along with 
the proper calculation of opacities, including evaporation, have small but noticeable effects on
the thermal histories of chondrules.
Especially because of the post-shock radiation boundary condition, the shock speed needed for chondrule
melting increases slightly, from roughly $7 \, {\rm km} \, {\rm s}^{-1}$ (DC02), to 
$8 \, {\rm km} \, {\rm s}^{-1}$. 
Another difference, due to dust evaporation in the pre-shock region, is that the chondrules are exposed 
to high temperatures a longer period of time before they are melted ($\sim$ 4 hours, as opposed to $<$ 30 minutes). 

Significantly, we have found that molecular line cooling due to \wwater does {\it not} have a significant 
effect on the cooling rates of chondrules, contrary to the calculations of INSN and Miura \& Nakamoto (2006)
and, indeed, the estimates made by Morris et al.\ (2009). 
Primarily this is due to the effect we call backwarming. 
The warm ($\approx 1750 \, {\rm K}$) gas in the pre-shock region, and especially the hot ($\approx 2250 \, {\rm K}$)
gas immediately past the shock, radiate into the post-shock gas, heating it.  
This prevents the gas from cooling by line radiation until it has traveled to distances that are effectively
optically thick to line radiation.

A few key chondrule constraints remain mysterious in the context of the 1-D shock model (such as the lack of isotopic fractionation and the discovery of primary Na) and demand 2-D simulations.
Although the shock model for chondrule formation certainly needs further refinement, we have presented a model 
which corrects the major problems identified by Desch et al.\ (2005), representing the most complete 1-D model
of chondrule formation in nebular shocks. 
Most important, we have included a complete treatment of the molecular line cooling due to water.  
This effect had been predicted to greatly increase the cooling rates of chondrules, by more than 
$10^{4} \, {\rm K} \, {\rm hr}^{-1}$.
In contrast, our detailed study shows that line cooling plays a negligible role in chondrule cooling rates.  
  
Chondrule formation in nebular shocks is consistent with the detailed thermal histories of chondrules, with
the nebular setting implied by chondrule-matrix complementarity, the repeatability of the chondrule-forming
process, the inferred age difference between CAIs and chondrules, the compound chondrule frequency, the correlation 
between compound chondrule frequency and textural type, and many other constrains on chondrule formation.
Future challenges remain, especially identifying the source of the shocks: gravitational instabilities or
planetesimal bow shocks.
However, the evidence is overwhelming that chondrules formed by melting in shock waves in the solar nebula
protoplanetary disk. 

\acknowledgements This work was supported by NASA Origins of Solar Systems grant NNG06GI65G.  We would like to sincerely thank Fred Ciesla for his comments and suggestions, and for sharing his expert insight into the chondrule formation process.  We would also like to thank the anonymous referee for their insightful and useful remarks.


\appendix

\section{Jump Conditions}

Jump conditions relate physical conditions (e.g., density $\rho$, pressure $P$, temperature $T$,
velocity $V$) at a point before the shock to those after the shock. 
Immediately before and after the shock (i.e., a few meters), the jump conditions are those
of a classical ``adiabatic" shock because insignificant energy is radiated in that interval, 
and the role of solids can be neglected.
The compression of the gas is then $\rho_{2} / \rho_{1} = \eta_{\rm AD}^{-1}$, where 
\begin{equation}
\eta_{\rm AD} = \frac{ 2 \gamma }{ \gamma + 1 } \, \frac{1}{ \gamma {\rm M}^{2} } + 
                \frac{ \gamma-1 }{ \gamma+1 }
\end{equation}
(Mihalas \& Mihalas 1984).
Far from the shock, one must inlcude radiative fluxes and the effects of solids in the 
equations of mass, momentum and energy conservation.
We have done so and have developed new jump conditions appropriate far from the shock.  
As is standard, brackets refer in the following equations to a difference in the bracketed 
quantity between two positions; in this case, they are far ($\sim 10^{5} \, {\rm km}$) distances 
before and after the shock.  
Conservation of mass yields
\begin{equation}
\left[ 
\rho_{\rm g} V_{\rm g} + \rho_{\rm c} V_{\rm c}
\right] = 0.
\label{eq:mjump}
\end{equation}  
Conservation of momentum yields
\begin{equation}
\left[ 
\rho_{\rm g} V_{\rm g}^{2} + P_{\rm g} + \rho_{\rm c} V_{\rm c}^{2}
\right] = 0.
\label{eq:momjump}
\end{equation} 
Conservation of energy yields 
\begin{equation}
\left[
\rho_{\rm g} V_{\rm g} \, \left( \frac{1}{2} V_{\rm g}^{2}
+ \frac{\gamma}{\gamma - 1} \frac{P_{\rm g}}{\rho_{\rm g}} \right) 
+ \rho_{\rm c} V_{\rm c} \, \left( \frac{1}{2} V_{\rm c}^{2} +
 C_{\rm P} T_{\rm c} \right) + F_{\rm rad} 
\right] = 0.
\label{eq:ejump}
\end{equation} 
We now define
\begin{equation}
\frac{\gamma}{\gamma-1}P_{\rm g} + \rho_{\rm c} C_{\rm p} T_{\rm c}
\equiv \frac{\gamma'}{\gamma'-1}P_{\rm g}.
\end{equation}
Because solids are in dynamical equilibrium with the gas far from the shock, we define
$V_g = V_c \equiv V$, so 
\begin{equation}
\frac{\gamma'}{\gamma'-1} = \frac{\gamma}{\gamma-1}+\delta,
\end{equation}
where
\begin{equation}
\delta = (\rho_{\rm c} C_{\rm p} T_{\rm c})_0/P_0.
\end{equation}
This results in
a new effective adiabatic index
\begin{equation}
\gamma'= \frac{\gamma+\delta\left(\gamma-1\right)}{1+\delta\left(\gamma-1\right)},
\end{equation}
As $\delta \rightarrow 0, \gamma' = \gamma$ (recovering the pure gas case), 
and if $\delta \gg 1, \gamma' =1$ (i.e., the chondrule fluid has no pressure).
We can now rewrite equation~\ref{eq:ejump} as
\begin{equation}
\left[
\left( \rho_{\rm g} V + \rho_{\rm c} V \right)
\frac{1}{2} V^{2}
+ \frac{\gamma'}{\gamma' - 1}P_{\rm g}V + F_{\rm rad} 
\right] = 0.
\label{eq:ejump2}
\end{equation} 
We assume that far from the shock dynamical, thermal, and chemical equilibria are achieved,
and that all hydrogen is molecular.
Well before the shock, all components have identical velocity $V_{0}$
and temperature $T_{0}$; well after the shock, they have identical 
velocity $V_{\rm f}$ and temperature $T_{\rm f}$.
By the equations of continuity, $n_{\rm H2} V$ and $n_{\rm He} V$ are 
conserved.
We can therefore rewrite the jump conditions as
\begin{equation}
\left[ \left( \rho_{\rm g} V_{\rm g} \right) + 
       \left( \rho_{\rm c} V_{\rm c} \right) \right]_0 \, \left[ V_{0} - V_{\rm f} \right]
= P_{0} \, \left[ \left( \frac{ V_{0} }{ V_{\rm f} } \right) \, 
                  \left( \frac{    T_{\rm g,f} }{    T_{\rm g,0} } \right) - 1 \right], 
\end{equation}
and
\[
\left[ \rho_{\rm g} V_{\rm g} +  \rho_{\rm c} V_{\rm c} \right]_0
\left[ \frac{1}{2}V_{0}^2 - \frac{1}{2}V_{\rm f}^2 \right]
+ \frac{\gamma'}{\gamma'-1}P_{0}V_{0}
\left(1-\frac{P_{\rm f}V_{\rm f}}{P_{\rm 0}V_{\rm 0}} \right)
\]
\begin{equation}
= F_{\rm rad}(\tau = \tau_{\rm m}) - F_{\rm rad}(\tau = 0) = \Delta F.
\end{equation}Note that when written in this format, all terms on both sides of the equations 
are positive: 
$F_{\rm rad}(\tau = 0) < 0$ and $F_{\rm rad}(\tau = \tau_{\rm m}) > 0$, because radiation
is emitted from the region near the shock front, and $V_{0} > V_{\rm f}$ and $T_{\rm f} > T_{0}$.

We now simplify the equations using the following dimensionless quantities:
\begin{eqnarray}
\eta & = & \frac{ V_{\rm f} }{ V_{0} } < 1 \\
y & = & \frac{ T_{\rm f} }{ T_{0} } > 1 \\
\gamma \rm M^2& = & \frac{ \left( \rho_{\rm g} + \rho_{\rm c} \right) V_{0}^{2} }{ P_{0} }
\end{eqnarray}
This yields two equations for $\eta$ and $y$:
\begin{equation}
\left ( \gamma \rm M^2 \right ) \left( 1 - \eta^2 \right)
+ \frac{2 \gamma'}{\gamma'-1}\left(1-y \right)
= \frac{2 \Delta F}{P_0 V_0}.
\label{eq:prequad}
\end{equation}
and 
\begin{equation}
\gamma {\rm M}^{2} \, \left( 1 - \eta \right) = \frac{y}{\eta} - 1.
\end{equation}
Finally we arrive at the following quadratic for $\eta$: 
\begin{equation}
\left( 1 - \eta \right) \, \left( \eta'_{\rm AD} - \eta \right)
= \frac{1}{\gamma {\rm M}^{2}} \, \left( \frac{ \gamma' - 1 }{\gamma' + 1} \right) \, 
  \frac{2 \Delta F}{P_0 V_0} = \epsilon,
\label{eq:quad} 
\end{equation}
where 
\begin{equation}
\eta'_{\rm AD} \equiv \frac{\gamma'-1}{\gamma'+1}+\frac{2\gamma'}{\gamma'+1}\;\frac{1}{\gamma \rm M^2}
\end{equation}
and $\epsilon$ is the ratio of net outward radiative fluxes to kinetic energy flux, given by 
\begin{equation}
\epsilon = \frac{ \gamma'-1 }{ \gamma'+1 } \,
           \frac{ F_{2} - F_{1} }{ \rho_{1} V_{1}^{3} / 2 }.
\end{equation}
If radiation carries energy away from the shock front, the signs of $F_2$ and $F_1$ (or $\Delta F$)
guarantee $\epsilon > 0$.
(The neglect of the sign of the radiative fluxes is one of the flaws of the jump
conditions used by Hood \& Horanyi (1991) and DC02 as described in Desch et al. (2005)).
If $\Delta F =0$, i.e., there are no radiative losses, then $\eta = \eta'_{\rm AD}$,
reducing to the adiabatic case but including the energy carried by solids. 
For $\Delta F \neq 0$,
\begin{equation}
\eta = \frac{\eta'_{\rm AD}+1}{2}
- \frac{1-\eta'_{\rm AD}}{2} \left(1+\epsilon \right)^{1/2}.
\end{equation}
Once $\eta$ is found, the jump condition governing the compression of the gas is 
$\rho_{2} / \rho_{1} = \eta^{-1}$, and the post-shock temperature is 
\begin{equation}
T_{\rm post} = T_{\rm pre} \, \eta \, [ 1 + \gamma {\rm M}^{2} (1 - \eta) ].
\label{3}
\end{equation}

\section{Continuum Radiative Transfer}

Solids are allowed to absorb and emit continuum radiation, which affects 
their energy budget.  Here we describe how we calculate the frequency-integrated
mean intensity of the radiation field at all locations. 

The first parameter to be defined is the (continuum) optical depth at all locations.
At the post-shock boundary, the optical depth $\tau(x=+X_{\rm post}) = 0$,
increasing to a maximum value $\tau(x=-X_{\rm pre}) = \tau_{\rm m}$ at the
pre-shock boundary. 
At other locations, 
\begin{equation}
\tau (x) = \int_{x'=x}^{x=+X_{\rm post}} \left[ \rho_{\rm g} \kappa + 
 \sum_{j=1}^{J} n_{j} \pi a_{j}^{2} \epsilon_{j} 
 \right] \, {\rm d} x',
\label{eq:tau} 
\end{equation}
where $\kappa$ is the opacity of the gas with density $\rho_{\rm g}$, due 
to dust associated with it.
If at some point the gas is so hot and dense that dust is destroyed (\S 3.1),
then $\kappa$ is set to zero from that point on.

The second parameter to be defined is the source function, $S$.
The source function from a blackbody at temperature $T$ is the Planck 
function $B_{\lambda}$, which after integrating over wavelength is $B = \sigma T^{4}$,
where $\sigma$ is the Stefan-Boltzmann constant (our radiative terms 
$S$, $B$ and $J$ are a factor of $\pi$ greater than their usual definitions, 
for ease of notation).
If all the particles in a region are held at temperature $T$, the source function 
must approach $B$, regardless of the emissivities of the particles.
If the particles are at different temperatures, the source function is weighted
according to their emissivities: 
\begin{equation} 
S = \frac{ \rho_{\rm g} \kappa \sigma T_{\rm g}^{4} + 
            \sum_{j=1}^{J} n_{j} \pi a_{j}^{2} \epsilon_{j} \, \sigma T_{j}^{4} }
      { \rho_{\rm g} \kappa + \sum_{j=1}^{J} n_{j} \pi a_{j}^{2} Q_{j} }.
\end{equation}
where $Q_j$ is the (wavelength-integrated) absorption coefficient, and the index $j$ 
represents potential multiple populations of solids. 
This reduces to $\sigma T_{\rm g}^{4}$ in the event that all $T_{j} = T_{\rm g}$ 
are identical.
(Although the temperature of dust is approximated in the code, the gas 
temperature is used for the source function from dust to avoid numerical
instabilities, and because the two are so similar.)
The last parameter to be specified is the radiation entering through the
two computational boundaries (integrated over wavelength).
The radiation field entering the pre-shock computational boundary is given by
$I_{\rm pre} = \sigma T_{0}^{4}$, where $T_{0}$ is the temperature of the ambient 
medium.
The radiation field entering the post-shock boundary is given by 
$I_{\rm post} = \sigma T_{\rm post}^{4}$, where $T_{\rm post}$ is the 
post-shock equilibrium temperature, using the radiative jump conditions
of Appendix A.

Given the incident radiation fields and the source function at all optical 
depths, $S(\tau)$, the mean intensity of radiation, $J(\tau)$ (integrated over 
wavelength) can be found:
\begin{equation}
J(\tau) = \frac{I_{\rm pre }}{2} {\rm E}_{2} ( \tau_{\rm m} - \tau )
         +\frac{I_{\rm post}}{2} {\rm E}_{2} ( \tau ) + \frac{1}{2}
         \int_{0}^{\tau_{\rm m}} S(t) {\rm E}_{1} \left| t - 
         \tau \right| \, {\rm d} t,
\end{equation}
where ${\rm E}_{1}$ and ${\rm E}_{2}$ are exponential integrals.
Using the properties of the exponential integrals, namely that 
${\rm d} {\rm E}_{n}(x) / {\rm d} x = - {\rm E}_{n-1} (x)$, it
can be shown by direct integration that if $S(t) = I_{\rm pre} =
I_{\rm post} = I_{0}$ everywhere, then $J(\tau) = I_{0}$.
We can also solve for the net flux of radiation energy, $F_{\rm rad}(\tau)$:
\[
F_{\rm rad}(\tau) = 
  + 2 I_{\rm pre } {\rm E}_{3} ( \tau_{\rm m} - \tau )
  + 2 \int_{\tau}^{\tau_{\rm m}} S(t) {\rm E}_{2} ( t - \tau ) \, {\rm d} t
\]
\begin{equation}
  - 2 I_{\rm post} \, {\rm E}_{3} ( \tau )
  - 2 \int_{0}^{\tau} S(t) {\rm E}_{2} (\tau - t ) \, {\rm d} t.
\end{equation}
It is again possible to show by direct integration that in the case
$S(t) = I_{\rm pre} = I_{\rm post} = I_{0}$ everywhere that the net flux 
$F_{\rm rad} = 0$.
It is also straightforward to demonstrate that in general 
\begin{equation}
\frac{ \partial F_{\rm rad} }{ \partial x } = 
 -4 \rho_{\rm g} \kappa \left[ J_{r} - \sigma T_{\rm g}^{4} \right]
 -\sum_{j=1}^{J} n_{j} 4\pi a_{j}^{2} \, \epsilon_{j} 
                        \left[ J_{r} - \sigma T_{j}^{4} \right],
\label{eq:dfdx}
\end{equation} 
where $\kappa$ is the opacity of the gas (via the well-coupled dust
component).

\section{Line Radiation Transfer}

\begin{figure}[ht]
\includegraphics[width=14cm]{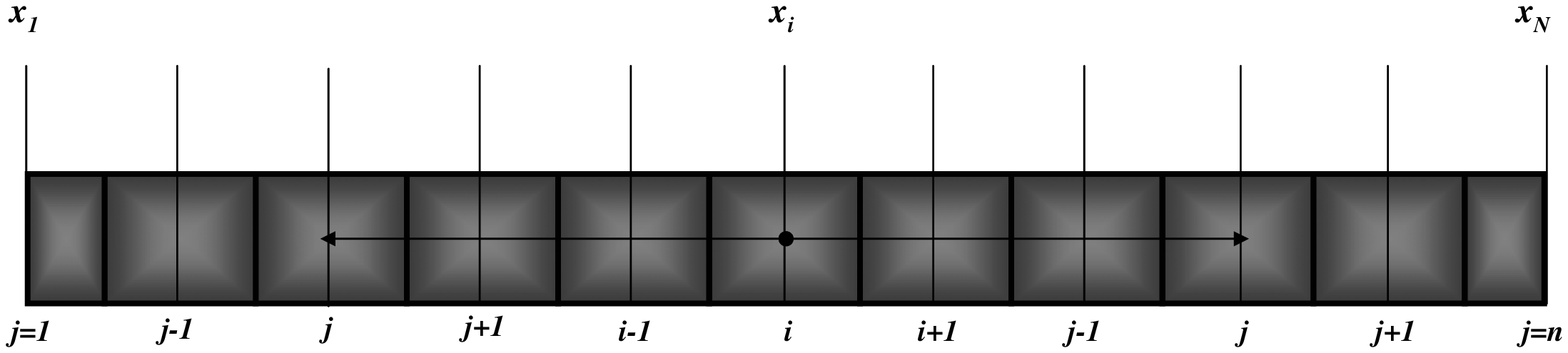}
\caption{The transfer of line radiation from zone $i$ to zone $j$.}
\label{fig:grid}
\end{figure}

In addition to continuum radiation, line radiation is emitted by ${\rm H}_{2}{\rm O}$
molecules.  
This radiation can be absorbed by other ${\rm H}_{2}{\rm O}$ molecules, or by solids.
${\rm H}_{2}{\rm O}$ molecules can also absorb line radiation or continuum radiation. 
We calculate the transfer of line radiation using $\dot{E}_{i \rightarrow j}$, the change
in energy per time (per area), due to line emission {\it from} the zone at index $i$, {\it into} 
a different zone $j \neq i$ (see Figure~\ref{fig:grid}).
This rate depends on the rates of line emission, $\Lambda$, and the density of water molecules,
$n_{\rm H_2O}$, in both zones, because water molecules
in $i$ can emit to water molecules in zone $j$, but they can also absorb the radiation from
molecules in zone $j$.  
The term $\dot{E}_{i \rightarrow j}$ also depends on the column density of water, $N_{\rm H_2O}$,
and the effective column density of solids, $\Sigma_{\rm eff}$ (as defined in \S 3.2), between the two 
zones. 

In our numerical code, gas quantities such as temperature, $T_{i}$, are stored at positions 
labeled $x_{i}$, where $i$ varies from 1 to $N$, $N$ assumed odd. 
The ``zone" with temperature $T_{i}$ is assumed to span the distance from
$x = (x_{i-1} + x_{i})/2$ to $x = (x_{i} + x_{i+1})/2$, except at the boundary $i=1$, where zone 1
spans $x = x_{1}$ to $(x_{1} + x_{2})/2$ and at the boundary $x = x_{N}$, where zone $N$ spans
$x = (x_{N-1} + x_{N})/2$ to $x_{N}$. 

Energy emitted by zone $i$ is assumed to be absorbed by another zone $j$ if line radiation can
propagate to zone $j$, but does not propagate into the next farthest zone ($j+1$ if $i < j$, or 
$j-1$ if $i > j$). 
This depends on the water column densities between the two zones, or equivalently the difference
in water column densities from the computational boundary (at $x = x_{1}$) to zone $i$ and to zone $j$,
and likewise on the difference in solids column density between the two zones. 
The energy from zone $i$ absorbed in zone $j$ is linearly proportional to the water density in zone $i$, 
$n_{\rm H_2O_i}$, and the width of zone $i$, $dx_{i} = (x_{i+1}-x_{i-1})/2$.
(It is understood that $dx_{1} = (x_{2}-x_{1})/2$ and $dx_{N} = (x_{N} - x_{N-1})/2$.)
For various combinations of zones $i$ and $j$, the formulas employed are as follows:

\noindent 
if $j=1$ (and $i > j$), then 
\[
\dot{E}_{i \rightarrow j} =  n_{\rm H_2O_i} \; \left[ \Lambda(N_{\rm H_2O_{i}}-N_{\rm H_2O_{j+1}},\Sigma_{{\rm eff}_{i}}-\Sigma_{{\rm eff}_{j+1}},T_i)
\right.
\]
\begin{equation}
\left.
- \Lambda(N_{\rm H_2O_{i}}-N_{\rm H_2O_{j}},\Sigma_{{\rm eff}_{i}}-\Sigma_{{\rm eff}_{j}},T_i) \right] \; dx_i;
\end{equation}

\noindent 
if $2 \leq j \leq i-1$ (so $i > j$), then
\[
\dot{E}_{i \rightarrow j} = \frac{1}{2} \; n_{\rm H_2O_i} \; \left[ \Lambda(N_{\rm H_2O_{i}}-N_{\rm H_2O_{j+1}},\Sigma_{{\rm eff}_{i}}-\Sigma_{{\rm eff}_{j+1}},T_i)
\right.
\]
\begin{equation}
\left.
- \Lambda(N_{\rm H_2O_{i}}-N_{\rm H_2O_{j-1}},\Sigma_{{\rm eff}_{i}}-\Sigma_{{\rm eff}_{j-1}},T_i) \right] \; dx_i;
\label{eq:edot2}
\end{equation}

\noindent 
if $i+1 \leq j \leq N-1$ (so $i < j$), then 
\[
\dot{E}_{i \rightarrow j} = \frac{1}{2} \; n_{\rm H_2O_i} \; \left[ \Lambda(N_{\rm H_2O_{j-1}}-N_{\rm H_2O_{i}},\Sigma_{{\rm eff}_{j-1}}-\Sigma_{{\rm eff}_{i}},T_i)
\right.
\]
\begin{equation}
\left.
- \Lambda(N_{\rm H_2O_{j+1}}-N_{\rm H_2O_{i}},\Sigma_{{\rm eff}_{j+1}}-\Sigma_{{\rm eff}_{i}},T_i) \right] \; dx_i;
\label{eq:edot4}
\end{equation}

\noindent 
and if $j = n$ (so $i < j$), then
\[
\dot{E}_{i \rightarrow j} = n_{\rm H_2O_i} \; \left[ \Lambda(N_{\rm H_2O_{j-1}}-N_{\rm H_2O_{i}},\Sigma_{{\rm eff}_{j-1}}-\Sigma_{{\rm eff}_{i}},T_i)
\right.
\]
\begin{equation}
\left.
- \Lambda(N_{\rm H_2O_{j}}-N_{\rm H_2O_{i}},\Sigma_{{\rm eff}_{j}}-\Sigma_{{\rm eff}_{i}},T_i) \right] \; dx_i.
\label{eq:edot3}
\end{equation}

Note that for each combination of zones $i$ and $j$, the temperature $T_{i}$ is used to calculate the 
line emission from $i$ into $j$.
This means when calculating the reverse term, $E_{j\rightarrow i}$, the temperature $T_{j}$ will be 
considered; if the temperatures in the two zones are identical, transfer of line radiation is possible
only if the water densities differ.
If the temperatures and water densities are identical in zones $i$ and $j$, there will be no transfer
of energy between them via line photons.
Significantly, water molecules in zone $i$ might be at high temperature and actively emitting 
line radiation, but they will not cool the region unless there is a cooler medium for that
line emission to radiate into.

One special circumstance that must be considered is the case where zones $i$ and $j$ lie on opposite
sides of the shock front. 
In that case, the water molecules in the two zones will be sufficiently Doppler shifted (by roughly
$6 \, {\rm km} \, {\rm s}^{-1}$, far greater than the post-shock sound speed
$\approx 1 - 2 \, {\rm km} \, {\rm s}^{-1}$), that photons emitted by zone $i$ will not be absorbed by
water molecules in zone $j$, and vice versa. 
To account for this, when calculating $E_{i \rightarrow j}$, we effectively set $n_{\rm H_2O_j} = 0$
in zones on the opposite side of the shock front, when calculating the water column densities. 
By doing so, $N_{\rm H_2O_{j+1}} = N_{\rm H_2O_{j-1}}$, and 
$\Lambda(N_{\rm H_2O_{j+1}}) - \Lambda(N_{\rm H_2O_{j-1}}) = 0$ in equation B2.
In this and the other equations, water molecules on the other side of the shock front from zone $i$
are thus seen not to absorb radiation from zone $i$.
Solids in zone $j$, having a continuum opacity, can absorb these line photons, though.
However, if there were no solids in zone $j$, then $E_{i \rightarrow j}$ would vanish.   

The net line cooling in zone $i$ is then given by
\begin{equation}
\dot{e}_i = \frac{1}{dx_i}\; \sum_{j \neq i} \; \left( \dot{E}_{j\rightarrow i} - \dot{E}_{i\rightarrow j} \right).
\label{eq:edotnet}
\end{equation}
It is this term that is then added to the gas energy equation (i.e., to the right-hand side of equation 23 
of DC02).
Except for the line radiation that crosses the computational boundaries, no energy emitted as line
photons is lost from the computation.
Energy is for practical purposes conserved by this approach.
(Some energy can be lost, but only within a column density $N_{\rm H2O} \sim 10^{20} \, {\rm cm}^{-2}$
of the boundaries, or $\sim 10^4$ km of the boundaries; but the input continuum radiation field
prevents the gas from cooling there anyway.)
An important corollary is that if there are zones that experience a net cooling, there must be other
zones (at lower temperature) that experience a net heating due to line radiation.

Our algorithm assumes there is no radiative exchange of energy between chondrules and water molecules
in the gas. 
Absorption of line photons by chondrules can be neglected only if they would contribute negigibly to the 
chondrule energy budget, and if the opacity of chondrules does not affect the transfer of line photons.
Line photons make up a negligible fraction of the total energy density of the radiation field:
\[
\frac{ u_{\rm line} }{ u_{\rm cont} } \sim 
\frac{ n_{\rm H_2O} \Lambda(T) }{4 \rho_{\rm g} \kappa \, \sigma T^{4}} \,
\frac{ l_{\rm mfp,line} }{ l_{\rm mfp, cont} },
\]
where $\kappa \approx 0.03 \, {\rm cm}^{2} \, {\rm g}^{-1}$ refers to the opacity of chondrules,
$\Lambda \approx 8 \times 10^{-12} \, {\rm erg} \, {\rm s}^{-1}$ is the optically thin line photon
emission rate, and $l_{\rm mfp}$ refer to the mean free paths of line or continuum photons.
Assuming line photons travel a typical water column density $\approx 10^{19} \, {\rm cm}^{-2}$,
$l_{\rm mfp,line} \approx 100 \, {\rm km}$ and $l_{\rm mfp,cont} \approx (\rho_{\rm g} \kappa)^{-1}$,
yielding  $u_{\rm line} / u_{\rm cont} \sim 0.02$. 
Chondrules are overwhelmingly heated by absorbing continuum radiation (and thermal exchange with the
gas), not by absorbing line photons: including absorption of line photons would increase our estimates
of chondrule temperatures by $< 10 \, {\rm K}$. 
Likewise, chondrules do not affect the transfer of line photons, since the mean free path of 
infrared photons through chondrules is $\approx 6 \times 10^{4} \, {\rm km}$, much greater than 
$l_{\rm mfp, line}$.
We also assume there is no absorption of continuum photons by water molecules. 
This is justified if these continuum photons do not affect the energy budget of the gas, and if absorption
of continuum photons by gas does not affect the transfer of the continuum radiation field.
While continuum photons overall dominate over line photons, photons with the specific wavelengths that 
water molecules can absorb are overwhelmingly produced by other gas molecules, not chondrules (as 
evidenced from the fact that at the wavelengths absorbed by water, $l_{\rm mfp,line} \ll l_{\rm mfp,cont}$).
Neglecting these contributions is justified.
Finally, while water molecules do absorb line photons effectively, 98\% of the continuum radiation 
field is unaffected by the presence of water molecules, and transfer of this component is calculated 
adequately in our code. 
Our approximations, which reduce the communication between the gas and chondrule fluids via radiation,
are therefore justified. 
At any rate, effective coupling between gas and chondrules does exist in the code, and including more 
radiative coupling would have reduced the ability of chondrules to cool.

\end{document}